\documentclass[10pt,epsf,letterpaper]{article}

\usepackage{amsmath}
\usepackage{amsfonts}
\usepackage{verbatim}
\usepackage{amssymb}
\usepackage{graphicx}
\usepackage{mathrsfs}
\usepackage{cite}
\usepackage{physics}
\usepackage{hyperref}
\usepackage[utf8]{inputenc}

\hypersetup{colorlinks=true,linkcolor=blue,	filecolor=magenta, 	urlcolor=blue,	citecolor=blue,	pdftitle={}}

\providecommand{\U}[1]{\protect\rule{.1in}{.1in}}

\def\pa{\partial}

\graphicspath{{./fig_crop/}}
\renewcommand{\(}{\left(}
\renewcommand{\)}{\right)}
\renewcommand{\[}{\left[}
\renewcommand{\]}{\right]}

\begin{document}


\title{On the existence of thermodynamically stable asymptotically flat black holes}

\author{Dumitru Astefanesei,$^{(1)}$ Romina Ballesteros,$^{(1,2)}$ Paulina Cabrera,$^{(1)}$ \\ Gonz\'alo Casanova$^{(1)}$ and Ra\'ul Rojas$^{(3)}$
\\\\\textit{$^{(1)}$Pontificia Universidad Cat\'olica de Valpara\'\i so, Instituto de F\'\i sica} \\\textit{Av. Brasil 2950, Valpara\'{\i}so, Chile}
\\\textit{$^{(2)}$Instituto de F\'{\i}sica Te\'orica UAM/CSIC,} \\\textit{	C/ Nicol\'as Cabrera, 13--15,  C.U.~Cantoblanco, E-28049 Madrid, Spain} 
\\\textit{$^{(3)}$Departamento de F\'isica, Universidad de Concepci\'on, Casilla 160-C, Concepci\'on, Chile}}

\maketitle
\begin{abstract}
We use the quasilocal formalism of Brown and York, supplemented with counterterms, to investigate the thermodynamics of asymptotically flat black holes. We consider two families of exact regular black hole solutions, which are thermodynamically stable. The first one consists of four-dimensional static charged hairy black holes  in extended supergravity. The second family consists of five-dimensional static charged black holes in Gauss-Bonnet (GB) gravity. Despite the fact that their characteristics are completely different, we found a striking similarity between their thermodynamic behaviour.
\end{abstract}

{\hypersetup{linkcolor=black}
\tableofcontents}



\section{Introduction}

The black hole represents the equilibrium end state of gravitational collapse and the relationship between thermodynamic entropy and the area of an event horizon is one of the most robust results in gravitational physics. Once quantum effects were taken into account, it was understood by Hawking \cite{Hawking:1975vcx} that black holes can emit radiation. While the Hawking radiation produces a very small effect that is not relevant from an empirical point of view, the black hole thermodynamics makes sense only in a theoretical framework in which the black holes are thermodynamically stable. However, this is not the case for black holes in flat spacetime, e.g. Schwarzschild, Reissner-Nordstr\"om (RN), and Kerr black holes are thermodynamically unstable.

One way to circumvent this situation is to enclose the asymptotically flat black hole  by a finite `box' \cite{York:1986it} such that the radiation can not escape to the asymptotic region. However, while this proposal is interesting from a theoretical point of view, it does not provide any concrete hint of physical situations where this kind of boundary conditions can appear naturally. A more general construction, for a theory with a non-trivial negative cosmological constant, was considered by Hawking and Page \cite{Hawking:1982dh}. They found that, indeed, there exists a general family of `large' black holes (with the horizon radius comparable with the radius of anti-de Sitter (AdS) spacetime) that are thermodynamically stable. In this case, the conformal boundary of AdS (where the boundary conditions should be imposed) plays the role of the `box'. However, the boundary conditions are changed to correspond to a spacetime with negative cosmological constant
and, while this is a very important result in the context of AdS-CFT duality \cite{Maldacena:1997re}, it can not be applied to asymptotically flat black holes.

Therefore, to obtain a well defined thermodynamic framework, the key point is the existence of asymptotically flat black holes, which are thermodynamically stable, without imposing artificially boundary conditions corresponding to a finite box. Elucidating this aspect could be relevant for understanding the existence of supermassive black holes that can not be formed by gravitational collapse. One can imagine a scenario where, in the early universe, small black holes exist in a suitable environment that makes them thermodynamically stable, e.g. surrounded by dark matter. Then, after a long time, they can grow up to become supermassive if there is enough matter around they can absorb.

In this work, we carefully investigate the existence of such an `environment' and present a detailed analysis of the thermodynamic stability by using the quasilocal formalism supplemented with boundary counterterms. Both Gauss-Bonnet-Maxwell and Einstein-Maxwell-dilaton theories can be interpreted as particular completions of Einstein-Maxwell gravity, and so the question of how these theories modify some concrete features of black holes  could be of significant interest. First, we consider a four-dimensional theory, with a scalar field and its self-interaction, in which there exist exact hairy black hole solutions. While this is a toy model for static black holes, it shows that, even in a simple theoretical set up, the hair can stabilize, dynamically and thermodynamically, the black holes (see, also, \cite{Astefanesei:2019mds,Astefanesei:2019qsg,Astefanesei:2020xvn}). It can very well be that more complicated models for dark matter can have a similar effect. Second, we consider the five-dimensional gravity when the Einstein-Hilbert action is supplemented by GB corrections. These models can be understood, for example, as effective theories from string theory where the higher derivative corrections are tree level, depending of the string tension (there are also quantum corrections depending of the string coupling). It was found long time ago that there exist exact static neutral black hole solutions and they are thermodynamically stable \cite{Myers:1988ze} (see also \cite{Bueno:2016lrh, {Bueno:2017qce}}).\footnote{There is another physical mechanism that could be important for the stability of black holes in theories with higher derivative terms, namely the scalarization and existence of the scalar condensates \cite{Astefanesei:2020qxk}.} We consider the generalization of RN black hole in GB theory, compute the conserved charges  within the quasilocal formalism of Brown and York \cite{Brown:1992br} that is self-consistent and so we do not need Wald formalism \cite{Wald:1993nt} to obtain the entropy. We find again that there exist thermodynamically stable charged static black holes.

The quasilocal formalism supplemented with counterterms in flat spacetime was used in \cite{Astefanesei:2005ad} to obtain the thermodynamics of the dipole black ring \cite{Emparan:2004wy}. However, this formalism was put on a firm ground by Mann and Marolf in \cite{Mann:2005yr}, where they have shown that the addition of an appropriate covariant boundary term
to the gravitational action yields a well-defined variational principle for asymptotically
flat spacetimes. This becomes the standard tool that leads to a natural definition of conserved quantities at spatial infinity, e.g. \cite{Astefanesei:2006zd,Astefanesei:2009wi,Compere:2011db,Compere:2011ve,Astefanesei:2010bm, {Mann:2006bd}, {Mann:2008ay}}. In this work, we use  concrete counterterms to circumvent the problem with the background subtraction method when the `background' can not be explicitly constructed. We  construct the counterterm for GB gravity (and fix the ambiguity of an overall numerical factor), which is consistent with a correct variational principle and regularizes the Euclidean action. Interestingly, unlike the AdS case, there is no need of adding counterterms for the scalar field in flat spacetime when it vanishes at the boundary. We shall consider both the grand canonical ensemble, in which the electric potential difference between the event horizon and
the infinity is fixed, and the canonical ensemble, in which the electric charge is fixed.

The remainder of the paper is organized as follows: in Section \ref{sec2}, we present the exact asymptotically flat hairy charged black hole solution and study in detail the local and global thermodynamic stability. We use the counterterm method to obtain the on-shell regularized Euclidean action and the  quasilocal formalism to obtain the conserved charges. We identify the region of the phase space where there exist thermodynamically stable black holes. In Section \ref{sec3}, we present the charged black hole solution in GB gravity. We obtain the on-shell Euclidean action, conserved charges, Smarr formula, and verify the quantum-statistical relation.  Again, we identify the region of the phase space where there exist thermodynamically stable black holes.  In the last section, we present a brief review of our main results, complement the analysis of thermodynamic behaviour with an analysis using the thermodynamic potential, and discuss to some extent the extremal limit that is important for the existence of a consistent canonical ensemble. In Appendix \ref{localconditions}, we summarizes the general thermodynamic stability conditions of response functions for charged black holes. Finally, in Appendix \ref{SC}, we compare the counterterm method in AdS with the one in flat spacetime and explain why there is no need of scalar counterterms for regularizing the Euclidean action of asymptotically flat hairy black holes.	

\section{Black holes in Einstein-Maxwell-dilaton theory}
\label{sec2}

Let us consider the theory given by the action
\begin{equation}
	I=\frac{1}{2\kappa}\int_{\mathcal{M}}
	{d^4x\sqrt{-g}\[R -\frac{1}{2}(\pa\phi)^2- e^\phi{F^2}-V(\phi)\]}
	\label{action2}
\end{equation}
where $\kappa=8\pi$, in the unit system where $G=c=1$; $F^2=F_{\mu\nu}F^{\mu\nu}$, where $F_{\mu\nu}=\pa_\mu A_\nu-\pa_\nu A_\mu$ is the gauge field and $A_\mu$ the gauge potential, and $(\pa\phi)^2\equiv g^{\mu\nu}\pa_\mu\phi\pa_\nu\phi$.

\begin{figure}[t!]
	\centering
	\includegraphics[width=0.56\textwidth]{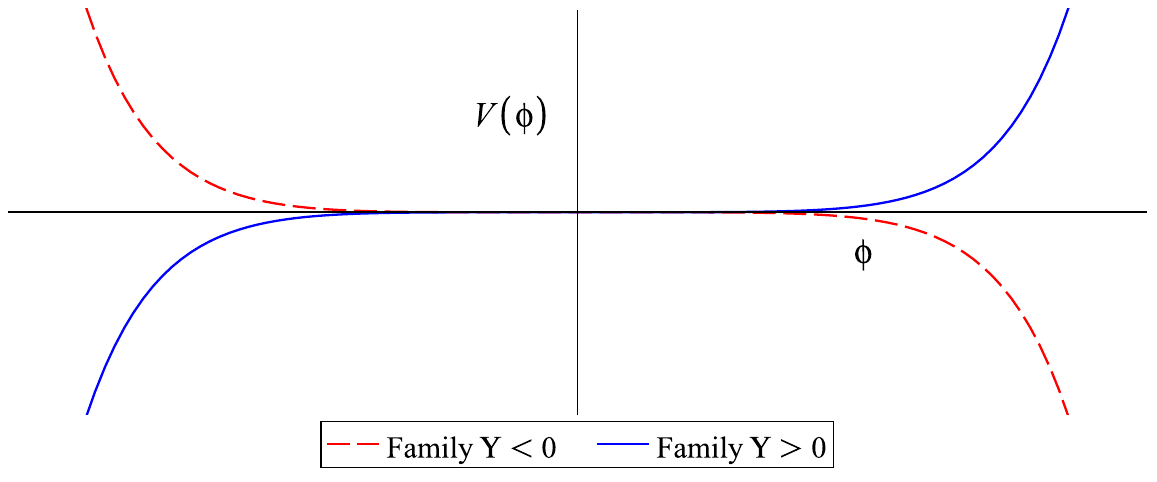} 
	\caption{Schematic behaviour of the scalar field potential for the family $\Upsilon<0$ and $\Upsilon>0$. The region $\phi<0$ corresponds to the negative branch and $\phi>0$ corresponds to the positive branch.}
	\label{figura0}
\end{figure}

The potential \cite{Anabalon:2012ta,Anabalon:2013eaa,Anabalon:2013qua} is 
\begin{equation}
	V(\phi)\equiv2\Upsilon\(2\phi+\phi\cosh\phi-3\sinh\phi\)
	\label{pot}
\end{equation}
where $\Upsilon$ is a dimensionful parameter that characterize the strength of the potential. Importantly, now it is well understood that in fact this is a model of a dilaton and its self-interaction, in the sense that the dilaton is endowed with a potential that originates from an 
electromagnetic Fayet-Iliopoulos (FI) term in $\mathcal{N} = 2$ extended supergravity in four spacetime dimensions \cite{Anabalon:2017yhv,Anabalon:2020pez} (see, also, \cite{Gallerati:2021cty}). Therefore, the theory we consider is consistent and its  ground state is stable (recently, exact hairy charged soliton solutions were constructed in \cite{Anabalon:2021tua, Anabalon:2022aig,Anabalon:2023oge, Anabalon:2024qhf}). The schematic behaviour of the potential (\ref{pot}) is depicted in Fig. \ref{figura0}. The potential is symmetric under $\phi\rightarrow-\phi$ and $\Upsilon\rightarrow-\Upsilon$ transformations. 

The exact static hairy black hole solution was obtained in \cite{Anabalon:2013qua},
\begin{equation}
	ds^2=\Omega(x)\[-f(x)dt^2+\frac{\eta^2dx^2}{x^2f(x)}+d\Sigma_2^2\]\,, \qquad A_\mu=\[\Phi-\frac{q(x-1)}{x}\]\delta_\mu^t\,, \qquad \phi(x)=\ln(x)
\end{equation}
where $d\Sigma_2^2\equiv d\theta^2+\sin^2\theta d\varphi^2$ and
\begin{equation}
	f(x)=\Upsilon\(\frac{x^2-1}{2x}-\ln{x}\) +\frac{\eta^2(x-1)^2}{x}\[1-\frac{2q^2(x-1)}{x}\]\,, \qquad
	\Omega(x)=\frac{x}{\eta^2(x-1)^2}
\end{equation}
In the expressions above, $x$ is a dimensionless coordinate that  is related to the standard canonical radial coordinate, $r$, by the transformation $\Omega(x)=r^2$. There are two constants of integration, $\eta$ and $q$, which are related to the mass and electric charge of the solution, and $\Phi$ in the expression of gauge field is an arbitrary additive constant. The black hole event horizon is located at $x=x_+$ such that $f(x_+)=0$ and $x_+$ is the biggest root. The asymptotic region is located at $x\rightarrow 1$, where the conformal factor $\Omega$ diverges, and the scalar field (along with its potential) vanishes.

We distinguish two families of solutions: the family for which $\Upsilon>0$ and the family for which $\Upsilon<0$. Within each family, there are two branches of solutions, characterized by the domain for the coordinate $x$, which in turn fixes the sign of $\phi$. These two branches are characterized by distinct boundary condition for the scalar field. To obtain the asymptotic behaviour of the dilaton, we consider the transformation to the canonical radial coordinate, namely $\Omega(x)=r^2$ near the boundary. This gives rise to two possible changes of coordinates,
\begin{equation}
	x=1\pm\frac{1}{\eta r}+\frac{1}{2\eta^2r^2}\pm\frac{1}{8\eta^3r^3}+\mathcal{O}(r^{-4})
\end{equation}
The negative branch corresponds to the case where $0\leq x<1$ (and so $\phi<0$). By using the corresponding change of coordinates, we obtain the following boundary expansion: $\phi=-\frac{\eta}{r}+\mathcal{O}(r^{-3})$. The positive branch corresponds to the case where $1<x\leq\infty$ (and so $\phi>0$), and the boundary expansion of the scalar field becomes $\phi=+\frac{\eta}{r}+\mathcal{O}(r^{-3})$.\footnote{Generally, for example in string theory where the scalar fields are moduli related to the coupling constants, the expansion of the scalar field in flat spacetime is $\phi=\phi_\infty + \Sigma/r +...$, where $\Sigma$ is the scalar charge. A discussion of why scalar charges (which are not conserved charges) \cite{Gibbons:1996af}, when $\phi_\infty$ can vary, do not appear in the first law of black hole thermodynamics was presented in \cite{Astefanesei:2018vga,Hajian:2016iyp}. However, since the theory we are interested in contains the dilaton potential, the boundary conditions for the dilaton are such that $\phi_\infty=0$ (see, also, \cite{Ballesteros:2023muf}).}

In the remaining of this section, we will be focusing on the family $\Upsilon>0$, which contains thermodynamically stable black holes. The family $\Upsilon<0$ only supports black hole solutions for the positive branch, but in this case the potential (\ref{pot}) becomes unbounded from below, as shown in Fig. \ref{figura0}.

\subsection{Euclidean action and thermodynamics}

Let us briefly present the computation of the on-shell Euclidean action. For concreteness, we perform the computations in the positive branch, that is, under the assumption that $x-1>0$. 

Both the bulk part of the action and the Gibbons-Hawking boundary term \cite{Gibbons:1976ue} are
\begin{align}
	\label{bulkGH}
	I^{E}_{bulk}+I_{GH}^{E}
	= \beta(-TS-\Phi Q)
	+\frac{4\pi\beta}{\kappa}\left[\frac{2}{\eta(x-1)}
	+\frac{\Upsilon-12\eta^{2}q^2+3\eta^2}{3\eta^{3}}+\mathcal{O}(x-1)\right] 
\end{align}
where the temperature and entropy are defined as usual: the temperature is obtained by removing the conical singularity in the Euclidean section, and the entropy is one quarter of the area of the event horizon,
\begin{equation}
	T=-\frac{x_+}{4\pi\eta}\left.\frac{df(r)}{dr}\right|_{x=x_+}\,, \qquad S=\pi\Omega(x_+)
\end{equation}
and $Q$ and $\Phi$ are the electric charge and its conjugate potential. The electric charge, in terms of the constants of integration, is obtained by using the Gauss law
\begin{equation}\label{charg}
	Q=\frac{1}{4\pi}\oint_{s_\infty^2}{e^\phi *F}=\frac{q}{\eta}
\end{equation}
and the conjugate potential is defined as
\begin{equation}
	\Phi=A_t(x\rightarrow 1)-A_{t}(x_+)=\frac{q(x_+-1)}{x_+}
\end{equation}

Now, in order to remove the divergent contribution appearing in (\ref{bulkGH}), it is necessary to add a gravitational counterterm for asymptotically flat spacetime \cite{Lau:1999dp,Mann:1999pc,Kraus:1999di,Mann:2005yr,Astefanesei:2005ad}\footnote{Unlike AdS spacetime where counterterms for the scalar field should be also included, in our case there is no need of counterterms associated to the dilaton. We discuss this point in great detail in Appendix \ref{SC}.}
\begin{equation}
	I_{ct}=-\frac{1}{\kappa}\int_{\pa\mathcal{M}}
	{d^3x\sqrt{-h}\sqrt{2\mathcal{R}^{(3)}}}
\end{equation}
where $\mathcal{R}^{(3)}$ is the Ricci scalar on the hypersurface $x=const$. With this foliation of the spacetime, the boundary $\pa\mathcal{M}$ consists of the hypersurface $x=const$ in the limit $x\rightarrow 1$. 
In the Euclidean section, we have
\begin{eqnarray}
	\label{ctgrav}
	I_{ct}^{E}=\frac{4\pi\beta}{\kappa}
	\[-\frac{2}{\eta(x-1)}
	-\frac{\Upsilon-12\eta^{2}q^{2}+6\eta^{2}}{6\eta^{3}}
	+\mathcal{O}(x-1)\]
\end{eqnarray} 
The total action is therefore
\begin{equation}\label{total1}
	I^E=I^{E}_{bulk}+I_{GH}^{E}+I_{ct}^{E}=\beta(-TS-\Phi Q)+\beta\left(\frac{12\eta^{2}q^{2}-\Upsilon}{12\eta^{3}}\right)
\end{equation}

The last term in (\ref{total1}) is indeed the mass of the black hole, as follows from computing the conserved charges of the system. We compute this energy by using the Brown-York quasilocal formalism \cite{Brown:1992br}, with the quasilocal boundary stress tensor
\begin{equation}
	\tau_{ab}\equiv -\frac{2}{\sqrt{-h}}\frac{\delta I}{\delta h^{ab}}
\end{equation}
where $I=I_{bulk}+I_{GH}+I_{ct}$, and the index $a$ stands for the coordinates on the hypersurface $x=const$. For this case, the concrete expression for $\tau_{ab}$ is \cite{Astefanesei:2005ad}
\begin{equation}
	\tau_{ab}
	=\frac{1}{\kappa}\left[
	K_{ab}-h_{ab}K-\Psi
	\left(\mathcal{R}^{(3)}_{ab}-\mathcal{R}^{(3)}h_{ab}
	\right)-h_{ab}\Box \Psi+\Psi_{;ab}\right]\,, \qquad \Psi\equiv\(\frac{2}{\mathcal{R}^{(3)}}\)^\frac{1}{2}
\end{equation}
Now, according to the Brown-York formalism, the conserved charge associated to the isometry generated by the killing vector $\xi^a$ is
\begin{equation}
	\label{energy1}
	E=\oint_{s_\infty^2}
	{d^2\sigma\sqrt{\sigma}N^a\xi^b\tau_{ab}}
\end{equation}
where $N^a$ is the timelike unit normal to the hypersurface $x=const$ and $\sigma$ is the determinant of the metric with $x=const$ and $t=const$. The integration is performed in the limit $x\rightarrow 1$. For $\xi^a=\delta^a_t$ and the result is
\begin{equation}
	E=\frac{12\eta^2q^2-\Upsilon}{12\eta^3}
\end{equation}
that also reproduces the mass read off from the expansion, in the canonical coordinates, of $g_{tt}$ in the asymptotic region. Therefore, the Euclidean on-shell action satisfies the quantum-statistical relation
\begin{equation}
	\frac{I^E}{\beta}=E-TS-\Phi Q
\end{equation}

The computation made so far has assumed that the boundary condition for the gauge field is $\delta A_t|_{\pa\mathcal{M}}=0$. This implies that the conjugate potential, $\Phi$, is fixed. Thus, the thermodynamic potential obtained, namely, $\mathcal{G}\equiv E-TS-\Phi Q$, corresponds to the grand canonical ensemble. One can construct the canonical ensemble, in which $Q$ is fixed instead, by performing a Legendre transform in ($\Phi$, $Q$), which is equivalent to adding to the action the following boundary term
\begin{equation}
	I_A=\frac{2}{\kappa}\int_{\pa\mathcal{M}}
	{d^3x\sqrt{-h}\,n_\nu F^{\mu\nu}A_\nu}
\end{equation}
giving rise to the thermodynamic potential $\mathcal{F}\equiv E-TS$.

Now that we have all the thermodynamic quantities consistently computed, one can verify that the first law of black hole thermodynamics is satisfied,
\begin{equation}
	dE=TdS+\Phi dQ
\end{equation}

\subsection{Equation of state $\Phi=\Phi(Q,T)$}

\begin{figure}[t!]
	\centering
	\includegraphics[width=0.48\textwidth]{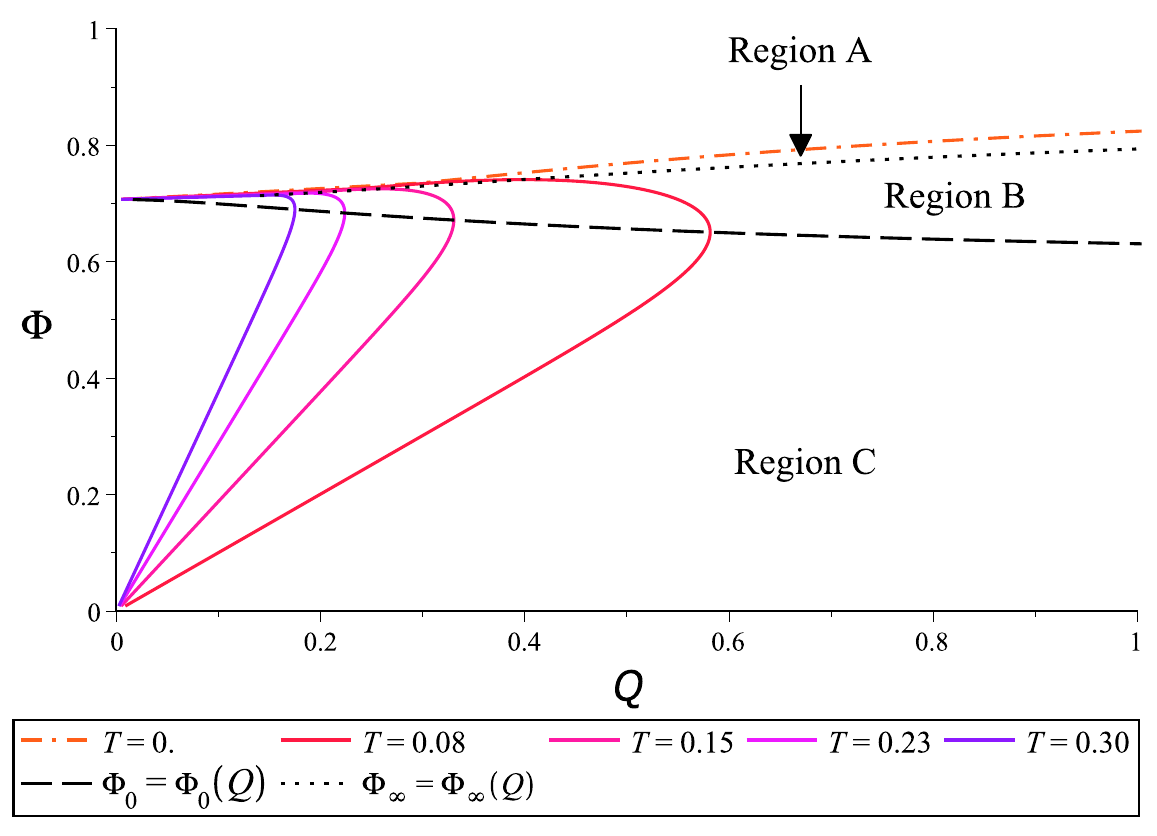} \quad
	\includegraphics[width=0.48\textwidth]{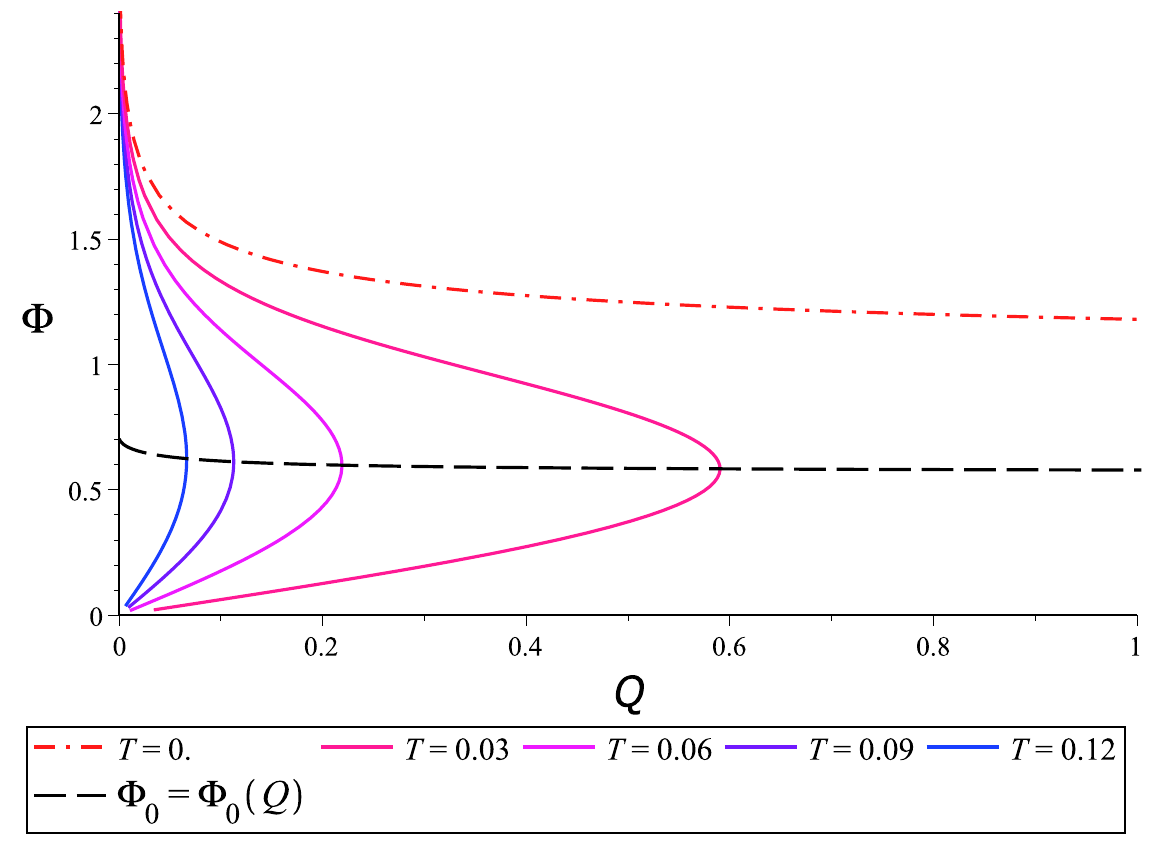} 
	\caption{\small Equation of state $\Phi-Q$, for $T$ fixed. \textbf{Left hand side:} Positive branch. \textbf{Right hand side:} Negative branch.}
	\label{EoS3}
\end{figure}

To proceed further with the thermodynamic behaviour, let us construct the equation of state $\Phi=\Phi(Q,T)$. First, we replace $q=Q\eta$ in the expressions for the thermodynamic quantities, as follows from (\ref{charg}). We then have the parametric expressions
\begin{equation}
	T={\frac{\left(x_{+}-1\right)\left[\left({\eta}^{4}{Q}^{2}-\frac{1}{2}{\eta}^{2}-\frac{1}{4}\Upsilon \right) {x_{+}^2}+ \left( {\eta}^{4}{Q}^{2}-\frac{1}{2}{\eta}^{2}+\frac{1}{4}\Upsilon\right)x_{+}-2{\eta}^{4}{Q}^{2} \right] }{2\pi\eta x_+^{2}}}\,, \quad \Phi=\frac{Q\eta(x_+-1)}{x_+}
\end{equation}
where $\eta$ must be isolated from the horizon equation $f(x_+)=0$. In Fig. \ref{EoS3}, we have depicted the equation of state for the positive and negative branch, respectively. In both cases we have considered the family $\Upsilon>0$. 

The equation of state contains relevant information on the local stability. Concretely, the response function
\begin{equation}
	\epsilon_T\equiv \(\frac{\pa Q}{\pa\Phi}\)_T
\end{equation}
known as isothermal permittivity is a measure of the stability of the configurations against small fluctuations of the electric charge (for more details, see \cite{Chamblin:1999tk,Chamblin:1999hg}). If $\epsilon_T>0$, the system is locally stable against electric fluctuations. In both cases, the positive and negative branches contain regions where $\epsilon_T>0$. For the negative branch (the plot at the right hand side of Fig. \ref{EoS3}), the region for which $\epsilon_T>0$ corresponds, as we are going to show in the next section, to the region where the second relevant response function, the heat capacity at constant electric charge, $C_Q$ (and also $C_\Phi$), is negatively defined, indicating a thermal instability. This is similar with  the thermodynamics of RN black hole and so these configurations are not locally stable.

The interesting case is for the positive branch (the plot on the left hand side of Fig. \ref{EoS3}), where a novel region with $\epsilon_T>0$ develops. This is the region A, as shown in the plot. It looks a very small region in the phase space, but it is valid for any $Q>0$ (and $T>0$). Next, we show that in this particular region, the heat capacity is also positive, thus fulfilling the conditions for local stability. The regions A, B and C are separated by the curves characterized by $(\pa\Phi/\pa Q)_T=0$ (the dotted curve, that separates A from B) and $(\pa Q/\pa\Phi)_T=0$ (the dashed curve, that separates B from C).

\subsection{Phase diagram and local stability}
The second response function relevant for the local stability is the heat capacity, both at fixed electric charge, $C_Q$, and at fixed conjugate potential $C_\Phi$, 
\begin{equation}
	C_Q\equiv T\(\frac{\pa S}{\pa T}\)_Q\,, \qquad 	C_\Phi\equiv T\(\frac{\pa S}{\pa T}\)_\Phi
\end{equation}
Thermally stable configurations are those with $C_Q>0$, for the canonical ensemble, and $C_\Phi>0$, for the grand canonical (see Appendix \ref{localconditions} for a brief summary on the criteria for local stability).

\subsubsection{Canonical ensemble: $Q$ fixed}
\begin{figure}[t!]
	\centering
	\includegraphics[width=0.47\textwidth]{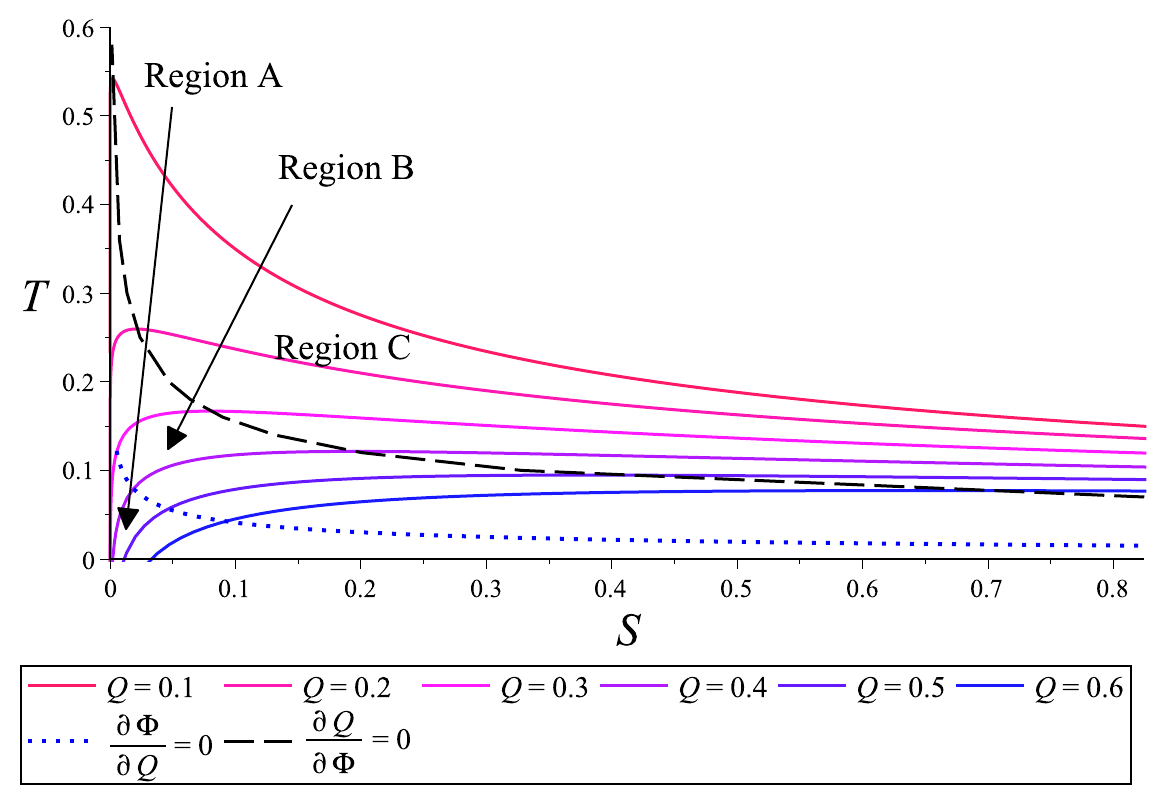}\quad
	\includegraphics[width=0.47\textwidth]{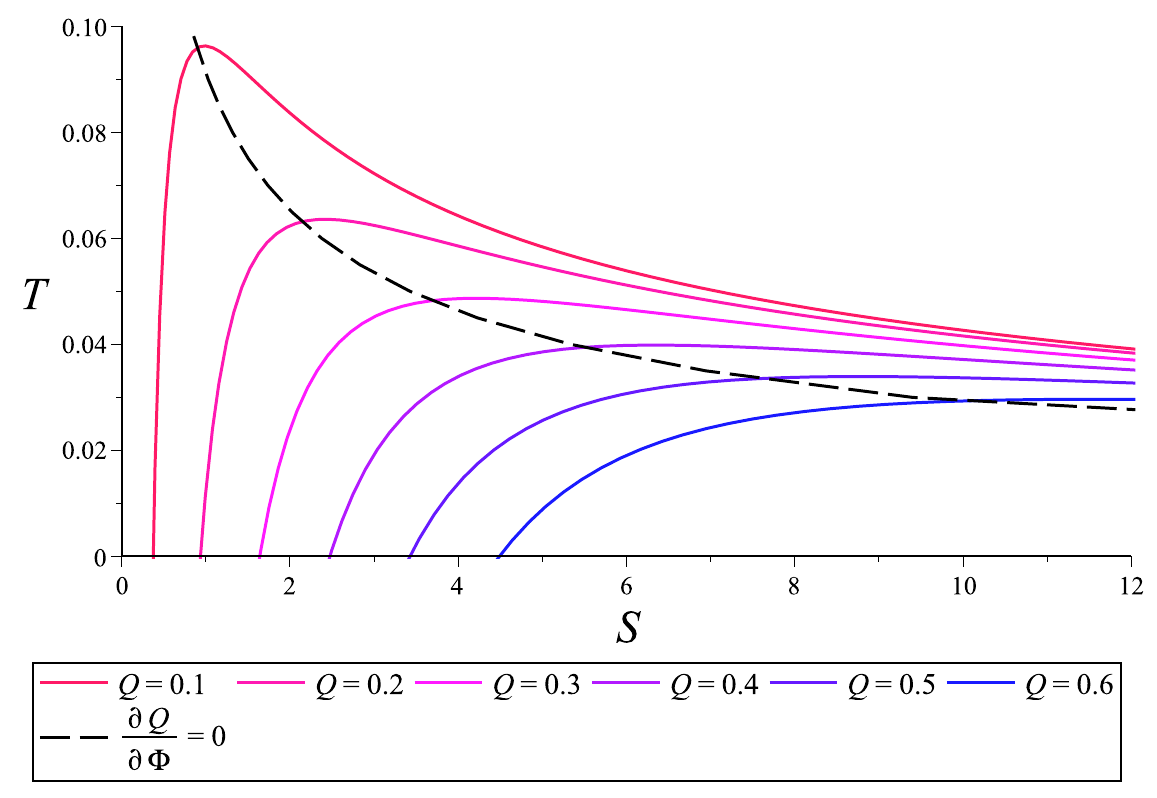} 
	\caption{\small $T-S$ for fixed $Q$. \textbf{Left hand side:} Positive branch. \textbf{Right hand side:} Negative branch.}
	\label{EoS5}
\end{figure}

In Fig. \ref{EoS5}, we have depicted the phase diagram $T-S$ of the canonical ensemble for the positive (left hand side plot) and negative branch (right hand side plot), respectively. As commented above, the interesting behaviour occurs for the positive branch, where we observe that $C_Q>0$ for both regions, A and B. However, only in region A we have, in addition, $\epsilon_T>0$ and, hence, only region A contains fully locally stable configurations, for the positive branch. On the other hand, the negative branch is characterized by two regions, according to the sign of $C_Q$. These two regions are separated by exactly the same curve that divides the regions $\epsilon_T>0$ and $\epsilon_T<0$ in the $\Phi-Q$ phase space (the dashed curve, for which $(\pa Q/\pa\Phi)_T=0$ and $C_Q^{-1}=0$) and, thus, the response functions $C_Q$ and $\epsilon_T$ can not be simultaneously positive for the negative branch.

\subsubsection{Grand canonical ensemble: $\Phi$ fixed}
\begin{figure}[t!]	
	\includegraphics[width=0.44\textwidth]{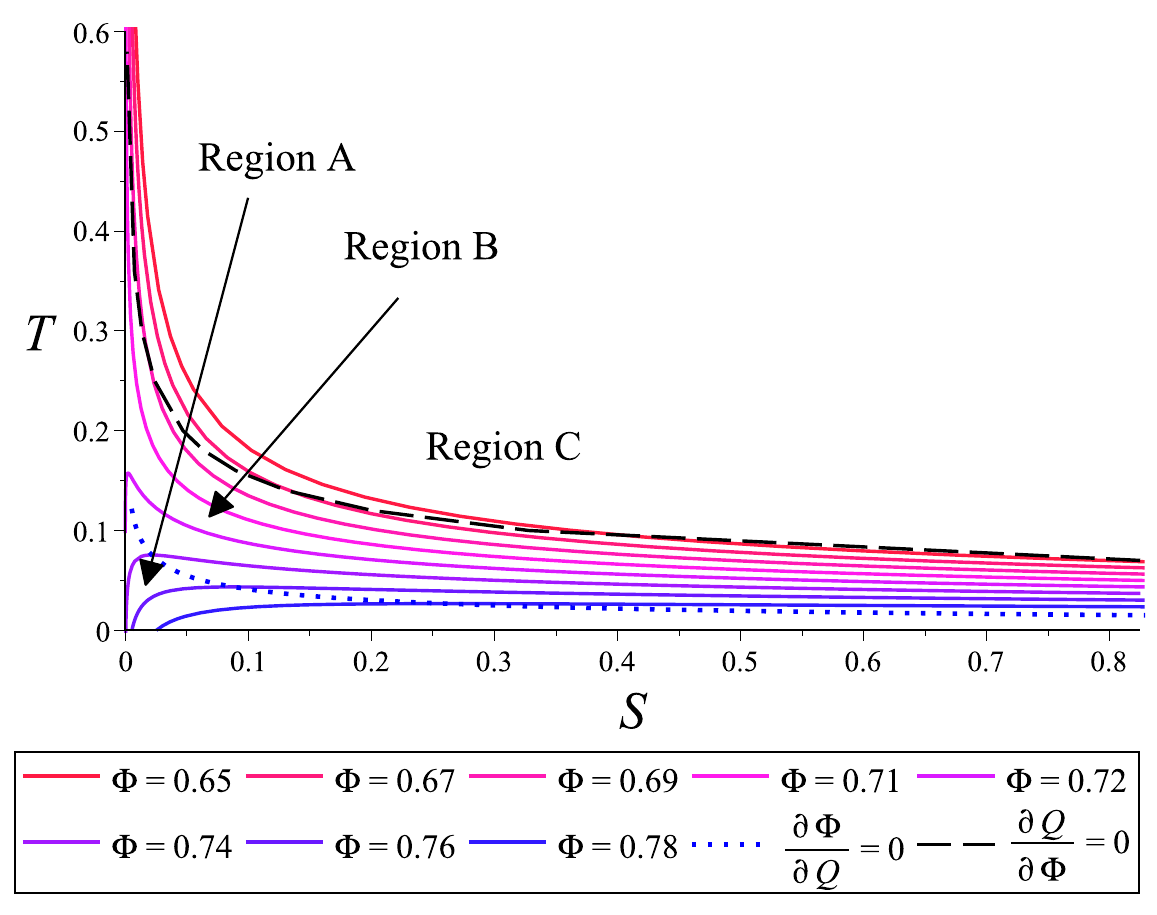} \quad
	\includegraphics[width=0.44\textwidth]{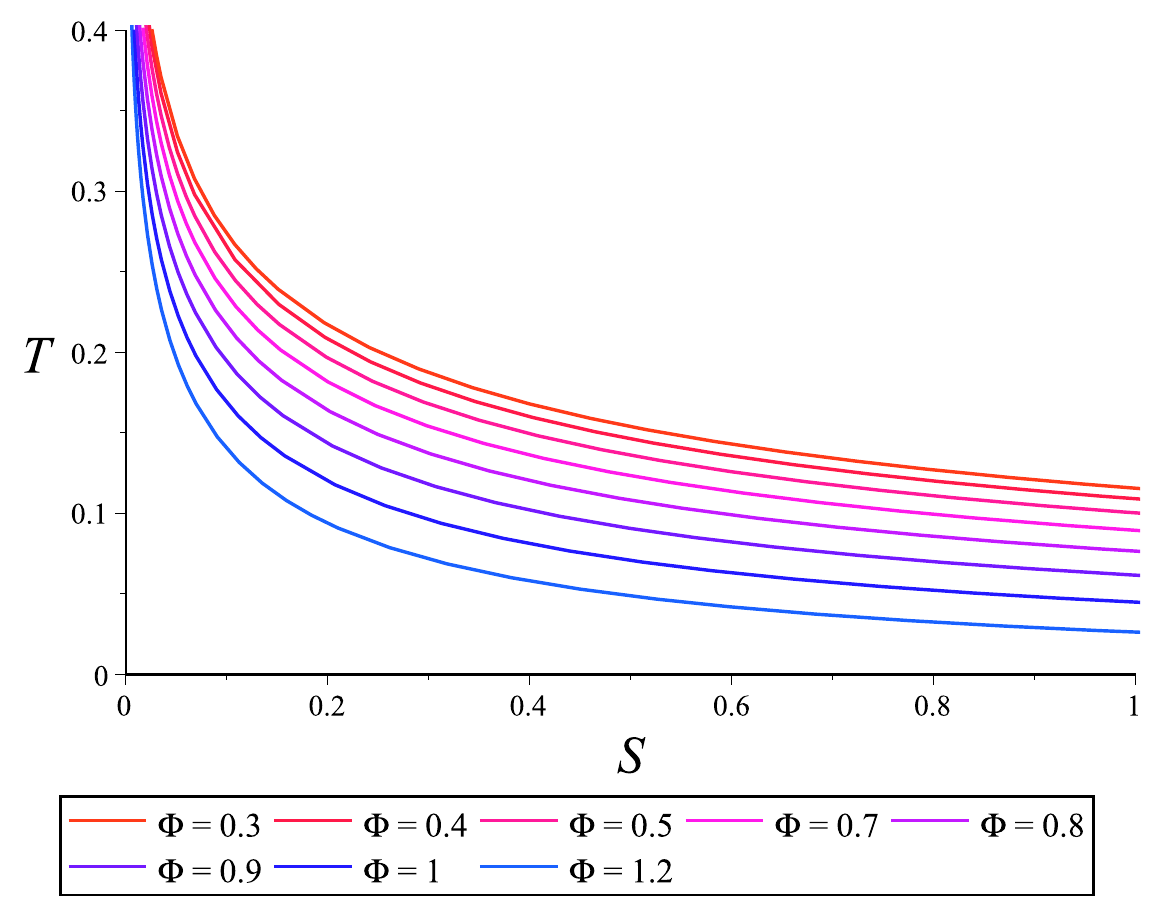} 
	\caption{\small $T-S$ for fixed $\Phi$. \textbf{Left hand side:} Positive branch. \textbf{Right hand side:} Negative branch.}
	\label{EoS4}\end{figure}

A similar analysis can be done in the grand canonical ensemble. As shown in Fig. \ref{EoS4}, since clearly $C_\Phi<0$, the black holes of the negative branch  are not stable. However, for the positive branch, we also observe that, in particular, the region A is characterized by $C_\Phi>0$, which also corresponds to $\epsilon_T>0$ and so there exist black hole solutions, which are thermodynamically stable (region B corresponds to $\epsilon_T<0$ and so the response functions are not simultaneously positively defined). We would also like to point out that, for the positive branch, the stable black holes are also globally stable, as they minimize the thermodynamic potential, $\mathcal{G}$. We are going to discuss this point in more detail in Discussion section.

\section{Black holes in Einstein-Maxwell-Gauss-Bonnet}
\label{sec3}

Let us now consider the Einstein-Maxwell-Gauss-Bonnet action \cite{Myers:1988ze}
\begin{equation}
	\label{action1}
	I=\frac{1}{2\kappa}\int_{\mathcal{M}}{d^5x\sqrt{-g}\(R+\frac{1}{4}\alpha\mathcal{R}_{GB}-F^2\)}
\end{equation}
where $\kappa=8\pi$, in the unit system where $G=c=1$, $F_{\mu\nu}=\pa_\mu A_{\nu}-\pa_\nu A_{\mu}$, $A_\mu$ is the gauge potential, $\mathcal{R}_{GB}\equiv R^2-4R_{\mu\nu}R^{\mu\nu}+R_{\mu\nu\sigma\rho}R^{\mu\nu\sigma\rho}$ is the GB  invariant and $\alpha$ is the coupling constant with GB sector. It is convenient to distinguish two branches, according to the sign of $\alpha$, since they have different thermodynamic properties, as shown below.

The equations of motion for this system are
\begin{equation}
	\label{motion}
	R_{\mu\nu}-\frac{1}{2}Rg_{\mu\nu}+\frac{1}{4}\alpha H_{\mu\nu}=\kappa T_{\mu\nu}^{EM}\,, \qquad
	\nabla_\mu F^{\mu\nu}=0
\end{equation}
where $T_{\mu\nu}^{EM}\equiv \frac{1}{4\pi}\(F_{\mu\alpha}F_{\nu}{}^{\alpha}-\frac{1}{4}g_{\mu\nu}F^2\)$ is the energy-momentum tensor for the electromagnetic field, and
\begin{equation}
	H_{\mu\nu}\equiv 2R_{\mu\alpha\beta\sigma}R_{\nu}{}^{\alpha\beta\sigma}-4R_{\mu\sigma\nu\rho}R^{\sigma\rho}-4R_{\mu\sigma}R_\nu^{\sigma}+2RR_{\mu\nu}-\frac{1}{2}\mathcal{R}_{GB}g_{\mu\nu}
\end{equation}
The spherically symmetric black hole solution was obtained in \cite{Wiltshire:1988uq}
\begin{equation}\label{ansatz}
	ds^2=-f(r)dt^2+\frac{dr^2}{f(r)} +r^2d\Sigma_{3}^2\,, \qquad A_\mu=\(\Phi-\frac{\sqrt{3}q}{2r^2}\)\delta_\mu^{t}
\end{equation}
where $d\Sigma^2_3\equiv d\theta^2+\sin^2\theta d\phi^2+\sin^2\theta\sin^2\phi d\psi^2$ is the line element of the unit 3-sphere, $\Phi$ is an additive constant in the expression of gauge potential, and $q$ is the charge parameter. Working with this ansatz, the differential equation for $f(r)$ becomes
\begin{equation}
	\(r^2+\alpha-\alpha f\)f'+2r\(f-1\)+\frac{2q^2}{r^3}=0
\end{equation}
and there exists two families characterized by $\epsilon=\pm 1$, which can be expressed in a compact form as
\begin{equation}\label{metric0}
	f(r)=1+\frac{r^2}{\alpha}\[1+\epsilon\({1+\frac{2\alpha\mu}{r^4}-\frac{2\alpha q^2}{r^6}}\)^\frac{1}{2}\]
\end{equation}
where $\mu$ is the constant of integration related to the mass. However, only the family $\epsilon=-1$ is consistent with asymptotic conditions of flat spacetime,  namely $f(r\rightarrow\infty)= 1+\mathcal{O}(r^{-2})$. The other family characterized by $\epsilon=1$  is not asymptotically flat, namely  $f(r\rightarrow\infty)= 2r^2/\alpha+1+\mathcal{O}(r^{-2})$. Since we are interested in asymptotically flat spacetime, in what follows we are going to consider the family $\epsilon=-1$.

As in general relativity, in GB theory the black hole configurations can also have at most two horizons for which $f(r)=0$,
\begin{equation}
	\label{horizon2}
	r_{\pm}=\frac{1}{2}\[2\mu-\alpha\pm \sqrt{(2\mu-\alpha)^2-16q^2}\]^\frac{1}{2}
\end{equation}
with the outer one, $r_+$, corresponding to the event horizon. We would like to emphasize that  the solution is regular when $2\mu-\alpha \geq 0$, otherwise it becomes a naked singularity. Therefore, the extremal black holes exist when the constraint  $(2\mu-\alpha)^2=16q^2$ is satisfied. In this case, the two horizons coincide ($r_+=r_-$) and so the mass parameter can be written as $\mu=2r_+ ^2+(1/2)\alpha$. 

In the next section, we shall explicitly compute  the regularized quasilocal stress tensor and energy. As a consistency check, we prove that, once an ambiguity in the overall factor that multiplies the counterterm is fixed, the quasilocal mass  matches, indeed, the Arnowitt-Deser-Misner (ADM) mass \cite{Arnowitt:1960es,Arnowitt:1960zzc,Arnowitt:1961zz,Arnowitt:1962hi}.

\subsection{Euclidean action and thermodynamics}

In this section, we use again the counterterm method and quasilocal formalism of Brown and York \cite{Brown:1992br} to consistently compute the regularized Euclidean on-shell action, boundary stress tensor, and energy. We verify that the quantum-statistical relation and first law of black hole thermodynamics are consistently satisfied. We also obtain Smarr formula in this non-trivial case.

The regularized action of this theory is
\begin{equation}
	\label{total}
	I^{E}=I^E_{{bulk}} +I^E_{GH}+I_{ct}^E
\end{equation}
where $I^E_{bulk}$ is the bulk part of the action given by (\ref{action1}), $I_{GH}$ and $I_{ct}$ are the Gibbons-Hawking boundary term \cite{Myers:1987yn} and the gravitational counterterm, respectively, with their corresponding extension for the GB sector, 
\begin{align}
	I^E_{GH}&=-\frac{1}{\kappa}\int_{\pa\mathcal{M}} {d^4x\sqrt{h^E}\(K+\frac{1}{2}\alpha\mathcal{K}_{GB}\)}
	\\I^E_{ct}&=\frac{1}{\kappa}\int_{\pa\mathcal{M}} d^4x\sqrt{h^E}\(\frac{3}{2}\mathcal{R}\)^\frac{1}{2}\(1+\frac{j}{9}\alpha\mathcal{R}\)
	\label{counter}
\end{align}
where $h$ is the determinant of the induced metric $h_{ab}$ on the boundary $\pa\mathcal{M}$, $K_{ab}=\nabla_a n_b$ is the extrinsic curvature, with $n_a$ being the normal unit vector to $\pa\mathcal{M}$, $K\equiv h^{ab}K_{ab}$, and 
\begin{equation}
	\mathcal{K}_{GB}\equiv \[\frac{2}{3}{K_{ac}K_{db}\(Kh^{cd}-K^{cd}\)}
	+\frac{1}{3}K_{ab}\(K_{cd}K^{cd}-K^{2}\)\]h^{ab} -2\(\mathcal{R}_{ab}-\frac{1}{2}\mathcal{R}h_{ab}\)K^{ab} 
\end{equation}
and $\mathcal{R}=h^{ab}\mathcal{R}_{ab}$ is the trace of the Ricci tensor on $\pa\mathcal{M}$.

For now, we use an arbitrary factor denoted by $j$ in the GB counterterm because the variational principle is well defined for any $j$. This ambiguity can be eliminated on physical grounds, and we are going to fix this factor in a consistent manner later on so that the physical quantities are well defined. A similar couterterm was used in \cite{Brihaye:2010wx} (see, also, \cite{Astefanesei:2008wz,Brihaye:2008xu} for GB counterterms in AdS).

The on-shell Euclidean action in the bulk can be written as\footnote{Here, $\beta=\int_0^\beta{d\tau}$ is the periodicity of the imaginary time in the Euclidean section.} 
\begin{equation}
	I^E_{\text{bulk}}=\frac{\pi\beta}{8}\left.\[r^3f'-{3\alpha}{r}\(f-1\)f'\]\right|_{r_+}^{r_b} -\frac{3\pi\beta q^2}{4r_+^2}
\end{equation}
where $f'\equiv df/dr$ and $r_b$ is the location of the boundary. We shall take the limit $r_b\rightarrow \infty$ at the end of the computation. Specifically, for RNGB black hole, the result for the bulk term can be written in the compact form as
\begin{equation}
	I^E_{\text{bulk}}=\frac{\pi\beta}{4}\(\mu-\frac{r_+^4+3\alpha r_+^2+2q^2}{r_+^2+\alpha}\)
\end{equation}
The Gibbons-Hawking boundary term is\footnote{Some intermediate results are \begin{equation}\mathcal{J}_{ab}h^{ab} =-\frac{f^\frac{1}{2}\(3rf'+2f\)}{r^3}\,, \;\; \mathcal{G}_{ab}{K}^{ab}=-\frac{3rf'+6f}{2r^3f^\frac{1}{2}}\notag\end{equation} where $\mathcal{G}_{ab}\equiv \mathcal{R}_{ab}-\frac{1}{2}\mathcal{R}h_{ab}.$}
\begin{equation}
	I_{GH}^E=\frac{\pi\beta}{2}(\mu-\alpha)-\frac{3\pi\beta}{4} r_b^2+\mathcal{O}\(r_b^{-2}\)
\end{equation}
Observe that $I_{GH}^E$ contains a divergent part $\propto{}r_b^2$, which does not come from the GB sector. As expected, this divergent quantity is going  to be removed with the gravitational counterterm that turns out to be
\begin{equation}
	I_{ct}^E=-\frac{\pi\beta}{2}\(\frac{3}{4}\mu-j\alpha\) +\frac{3\pi\beta}{4}r_b^2+\mathcal{O}\(r_b^{-2}\)
\end{equation}
All the divergent contributions are now cancelled out and the final result for the Euclidean action is
\begin{equation}
	\label{onshell}
	{I^E}=\frac{3\pi\beta}{8}\mu
	+\frac{\pi\beta}{2}(j-1)\alpha
	-\frac{\pi\beta}{4}\(\frac{r_+^4+3\alpha r_+^2+2q^2}{r_+^2+\alpha}\) 
\end{equation}
Notice that the second term in Eq. (\ref{onshell}), proportional to $\alpha$, is a finite contribution purely due to the GB sector. We shall see below that, for consistency with the asymptotically flat spacetime boundary conditions, we have to fix $j=1$ such that we recover the usual ADM mass.

We now compute the quasilocal energy for this system at spatial infinity. According to the Brown-York formalism,
\begin{equation}
	\label{energ}
	E_{quasi}=\int_{s_\infty^3}{d^3x\sqrt{\sigma}n^a\tau_{ab}\xi^{b}}
\end{equation}
is the conserved charge associated with the time symmetry of the metric, given by the Killing vector $\xi^a=\delta_t^a$, where $\sigma$ is the determinant of the metric on the 3-sphere $ds_{\sigma}^2=r^2d\Sigma_3^2$, $n_a=\delta_a^t/\sqrt{-g^{tt}}=f^\frac{1}{2}\delta_a^t$ is the normal unit to the hypersurface $t=$ constant, and $\tau_{ab}$ is the boundary stress tensor 
\begin{equation}
	\tau_{ab}\equiv\frac{2}{\sqrt{-h}}\frac{\delta I}{\delta h^{ab}}
	=\frac{1}{\kappa}\[K_{ab}-Kh_{ab} +\frac{1}{2}\alpha\(\mathcal{\tilde Q}_{ab}-\frac{1}{2}\mathcal{\tilde Q}h_{ab}\)\] -\frac{2}{\kappa}\tilde\Psi_{ab}
\end{equation}
where
\begin{equation}
	\mathcal{\tilde Q}_{ab}\equiv 3\mathcal{J}_{ab}+2K\mathcal{R}_{ab}-4\mathcal{R}_{ac}K^c_b+\mathcal{R}K_{ab}-2K^{cd}\mathcal{R}_{cadb}\,, \qquad \mathcal{\tilde Q}=\mathcal{\tilde Q}_{ab}h^{ab}
\end{equation}
and
\begin{equation}
	\tilde\Psi_{ab}\equiv \frac{d\Psi(\mathcal{R})}{d\mathcal{R}}\mathcal{R}_{ab}-\frac{1}{2}\Psi(\mathcal{R}) h_{ab}\,, \qquad \Psi(\mathcal{R})=\(\frac{3}{2}\mathcal{R}\)^\frac{1}{2} \[1+\frac{1}{9}{j}\alpha\mathcal{R}\]
\end{equation}

The components of the boundary stress tensor are
%
\begin{equation}
	\tau_{tt}=\frac{f}{8\pi r_b^3} \[3(r_b^2+\alpha)f^\frac{1}{2}-\alpha f^\frac{3}{2}-\(2j\alpha+3r_b^2\)\]=-\frac{1}{4\pi r_b^3}\[\frac{3\mu}{4}+(j-1)\alpha\]+\mathcal{O}(r_b^{-5})
\end{equation}
\begin{equation}
	\tau_{\theta\theta}=-\frac{4r_bf^\frac{1}{2}\(f^\frac{1}{2}-1\)+\(r_b^2+\alpha-f\alpha\)f'} {16\pi{}f^\frac{1}{2}} =-\frac{\mu^2-4q^2}{32\pi{}r_b^3} +\mathcal{O}(r_b^{-5})
\end{equation}
where $f=f(r_b)$ and $\tau_{\psi\psi}=\sin^2\phi\tau_{\phi\phi}=\sin^2\theta\sin^2\phi\tau_{\theta\theta}$. 

We have now all the necessary ingredients to compute the thermodynamic quantities. By using the formula (\ref{energ}), we get
\begin{equation}
	E_{quasi}=\frac{3}{8}\pi\mu+\frac{1}{2}\pi(j-1)\alpha
\end{equation}
This expression for the energy of RNGB black hole is consistent with the usual statistical formula obtained  from the thermodynamic potential  in the grand canonical ensemble:
\begin{equation}
	E=\left( \frac{\pa I^E}{\pa \beta} \right)_\Phi - \frac{\Phi}{\beta}\left( \frac{\pa I^E}{\pa \Phi} \right)_\beta=E_{quasi}
\end{equation}

A similar computation can be done in the canonical ensemble as $E=(\pa \tilde I^E/\pa\beta)_Q$, where \cite{Hawking:1995ap} 
\begin{equation}
	\tilde I=I+\frac{2}{\kappa}\int_{\pa\mathcal{M}}{d^4x\sqrt{-h}F^{\mu\nu}n_\mu A_\nu} \quad\rightarrow\quad \tilde I^E=I^E+\beta Q\Phi
\end{equation}
This action is compatible with the boundary condition for the gauge field that fixes the electric charge, which is given by the Gauss law,
\begin{equation}
	Q=\epsilon_0\oint{\star F}=\frac{\sqrt{3}\pi}{2}q
\end{equation}
and $\Phi\equiv A_t(\infty)-A_t(r_+)=\frac{\sqrt{3}}{2r_+^2}q$ is the conjugate potential. The Hawking temperature is computed as usual by removing the conical singularity in the Euclidean section,
\begin{equation}
	T=\beta^{-1}=\frac{1}{4\pi}f'(r_+) =\frac{r_+^4-q^2}{2\pi{}r_+^3(r_+^2+\alpha)}
	\label{temper}
\end{equation}
Before comparing the ADM and quasilocal masses, let us first proceed further with computing the entropy directly from the regularized action
\begin{equation}
	S= -I^E+\beta \left( \frac{\pa I^E}{\pa\beta}\right)_\Phi =\frac{1}{2}\pi^2r_+\(r_+^2+3\alpha\)
	\label{entro}
\end{equation}
that follows from statistical mechanics, after using the semi-classical approximation $\ln Z\approx e^{-I^E}$, where $Z$ is the partition function. The entropy receives a contribution from the GB sector and can be also computed by Wald formalism \cite{Wald:1993nt,Clunan:2004tb}. 

Before fixing $j$, let us first verify the quantum-statistical relation. There is some hope that we can eliminate the ambiguity related to $j$ by using this relation, but since the $\tau_{tt}$ component of the boundary stress tensor has a similar contribution, both contributions cancel each other in the quantum statistical relation. Concretely, we obtain\footnote{We have assumed the boundary condition $\delta A_\mu|_{\pa\mathcal{M}}=0$ for the gauge potential, that fixes $\Phi$ (the grand canonical ensemble). Similarly, $\mathcal{F}\equiv E-TS=\frac{3}{8}\pi\mu
	+\frac{1}{2}\pi(j-1)\alpha
	-\frac{1}{2}\pi^2r_+\(r_+^2+3\alpha\)=\beta^{-1}\tilde I^E$ for the canonical ensemble.} 
\begin{equation}
	\mathcal{G}\equiv E-TS-\Phi Q=\frac{3}{8}\pi\mu +\frac{1}{2}\pi(j-1)\alpha -\frac{1}{4}\pi\(\frac{r_+^4+3\alpha{}r_+^2+2q^2}{r_+^2+\alpha}\)=\beta^{-1}I^E
\end{equation}
and so it is satisfied for any $j$.

Therefore, the correct way to fix $j$ is to compare the quasilocal mass at spatial infinity with the ADM mass. Once $j$ is fixed for one specific solution, the counterterm can be used to regularize the energy of all regular  solutions with the same boundary conditions. We emphasize that the  fall off  of GB term is very fast at the boundary and that is why the asymptotic flat boundary conditions are permitted. 

Let us now prove that the quasilocal mass matches the ADM mass  only for $j=1$. We employ the ADM formalism as follows: first, we consider a generic equation of motion for GB gravity
\begin{equation}
	R_{\mu\nu}-\frac{1}{2}Rg_{\mu\nu}+\frac{1}{4}\alpha H_{\mu\nu}=\kappa T_{\mu\nu}
\end{equation}
It is convenient to rewrite it as
\begin{equation}\label{ADM1}
	R_{\mu\nu}+\frac{1}{4}\alpha\(H_{\mu\nu}-\frac{1}{3}Hg_{\mu\nu}\)=\kappa \bar T_{\mu\nu}
\end{equation}
where $H\equiv g^{\mu\nu}H_{\mu\nu}$ and $\bar T_{\mu\nu}=T_{\mu\nu}-\frac{1}{3}Tg_{\mu\nu}$, $T=g^{\mu\nu}T_{\mu\nu}$. Now, we expand the left hand side for a small perturbation, $g_{\mu\nu}=\mathring g_{\mu\nu}+h_{\mu\nu}$, where $|h_{\mu\nu}|\ll 1$ is the perturbation around a background spacetime given by $\mathring g_{\mu\nu}$. The left hand side of (\ref{ADM1}) expands as $-\frac{1}{2}\Box h_{\mu\nu}+\mathcal{O}(h^2)$ when the harmonic gauge condition is considered,
\begin{equation}
	\mathring\nabla_\mu\(h^{\mu\nu}-\frac{1}{2}\mathring g^{\mu\nu}h\)=0
\end{equation}
where $h=\mathring g^{\mu\nu}h_{\mu\nu}$. Since $|h_{\mu\nu}|\ll 1$, we consider the non-relativistic limit, under which $\Box h_{\mu\nu}\approx \nabla^2 h_{\mu\nu}$, $T\approx -T_{00}$. In this limit, we have the Poisson equation
\begin{equation}\label{poisson}
	\nabla^2 h_{\mu\nu}=-2\kappa\bar T_{\mu\nu}
\end{equation}
The solution to (\ref{poisson}) can be expressed as
\begin{equation}\label{hmunu}
	h_{\mu\nu}(x^i)=\frac{\kappa}{A}\int{\frac{\bar T_{\mu\nu}(y^i)d^4y}{|x-y|^2}}
\end{equation}
where $A=\int_0^\pi{\sin^2\theta d\theta}\int_0^\pi{\sin\phi d\phi}\int_0^{2\pi}{d\psi}=2\pi^2$ is the area of the unit 3-sphere. We obtain
\begin{equation}
	\nabla_x^2 h_{\mu\nu}(x^i)=\frac{\kappa}{A}\int{\bar T_{\mu\nu}(y^i)d^4y\nabla_x^2\(\frac{1}{|x-y|^2}\)}
\end{equation}
Notice that
\begin{equation}
	\int_V\nabla^2\(\frac{1}{r^2}\)dV=\oint_{S}{\nabla\(\frac{1}{r^2}\)dS}=-\frac{2}{r^3}\oint_S{dS}=-2A
\end{equation}
where $\oint_S dS=Ar^3$ is the area of the 3-sphere of radius $r$. Therefore,
\begin{equation}
	\nabla_x^2\(\frac{1}{|x-y|^2}\)=-2A\delta^4(x-y) \quad \rightarrow \quad	\nabla_x^2 h_{\mu\nu}(x^i)=-{2\kappa}\int{\bar T_{\mu\nu}(y^i)\delta^4(x-y)d^4y}=-2\kappa \bar T_{\mu\nu}(x^i)
\end{equation}
that is consistent with (\ref{hmunu}).

Now, to get the ADM mass, $M_{ADM}$, one must identify $M_{ADM}=\int{T_{00}d^4x}$. By expanding $h_{00}$ from (\ref{hmunu}) for the asymptotic region, $|x|\gg |y|$, we obtain
\begin{align}
	h_{00}(x^i)&=\frac{\kappa}{A}\[\frac{1}{r^2}\int{\bar T_{00}d^4y}+\mathcal{O}(r^{-4})\]\\
	&=\frac{\kappa}{A}\[\frac{1}{r^2}\int{\(T_{00}-\frac{1}{3}g_{00}T\)d^4y}+\mathcal{O}(r^{-4})\]\\
	&=\frac{\kappa}{A}\[\frac{2}{3r^2}\int{T_{00}d^4y}+\mathcal{O}(r^{-4})\]\\
	&=\frac{2\kappa M_{ADM}}{3Ar^2}+\mathcal{O}(r^{-4})
\end{align}
Therefore, the ADM mass can be directly read by obtaining $h_{00}$ from expanding $g_{00}$,
\begin{equation}
	M_{ADM}=\frac{3}{8}\pi r^2h_{00}
\end{equation}
where we have replaced $\kappa=8\pi$ and $A=2\pi^2$.
For the GB metric (\ref{metric0}), we asymptotically expand $g_{tt}$ to read $h_{00}$ around the flat spacetime,
\begin{equation}
	g_{00}=-1+\frac{\mu}{r^2}+\mathcal{O}(r^{-4})\,, \qquad h_{00}=\frac{\mu}{r^2}
\end{equation}

Therefore, only for $j=1$ we consistently obtain
\begin{equation}
	M_{ADM}=E_{quasi}=\frac{3}{8}\pi \mu
\end{equation}
and so this is the correct counterterm that is going to regularize the Euclidean action for any solution with the same boundary conditions.

We would also like to emphasize that in the limit $r_b\rightarrow\infty$, the trace of the boundary stress tensor vanishes, 
\begin{equation}
	\tau_{ab}h^{ab}=\frac{3\mu}{16\pi r_b^3} +\mathcal{O}(r_b^{-5})
\end{equation}
and it is covariantly conserved, $\tau^{a b}{}_{;b}=0$, that is compatible with our solution with no matter or conical defects at spatial infinity \cite{Astefanesei:2009mc}.

It is also straightforward to verify the first law of black hole thermodynamics,
\begin{equation}
	dE=TdS+\Phi dQ
\end{equation}
The Smarr formula is
\begin{equation}
	2E=3TS+2\Phi Q+2\mathcal{B}\alpha
\end{equation}
where the last term is, in principle, consistent with an extension of the first law, $dE=TdS+\Phi dQ+\mathcal{B}d\alpha$,
\begin{equation}
	\mathcal{B}\equiv\(\frac{\pa E}{\pa\alpha}\)_{S,Q} =-\frac{3}{8}\pi\(\frac{1}{2} +\frac{3r_+^2-2\mu}{r_+^2+\alpha}\)
\end{equation}
when the parameter $\alpha$ can vary, which is not our study case.


\subsection{Equation of state: $Q-\Phi$}

In this section, we obtain the equation of state and study in detail the regions where the system achieves local stability.

The equation of state $\Phi=\Phi(Q,T)$ can be implicitly written as
\begin{equation}
	T=\frac{1}{6}\(\frac{Q\Phi}{\pi}\)^\frac{1}{2} \frac{3-4\Phi^2}{\Phi\pi\alpha+Q}
\end{equation}
In the limit $\alpha=0$, the equation of state reduces to the one of charged black hole in general relativity. For this case, it is well known that there are no locally stable configurations, as $C_Q$ and $\epsilon_T$ can not be simultaneously positive. 

\begin{figure}[t!]
	\includegraphics[width=0.5\textwidth]{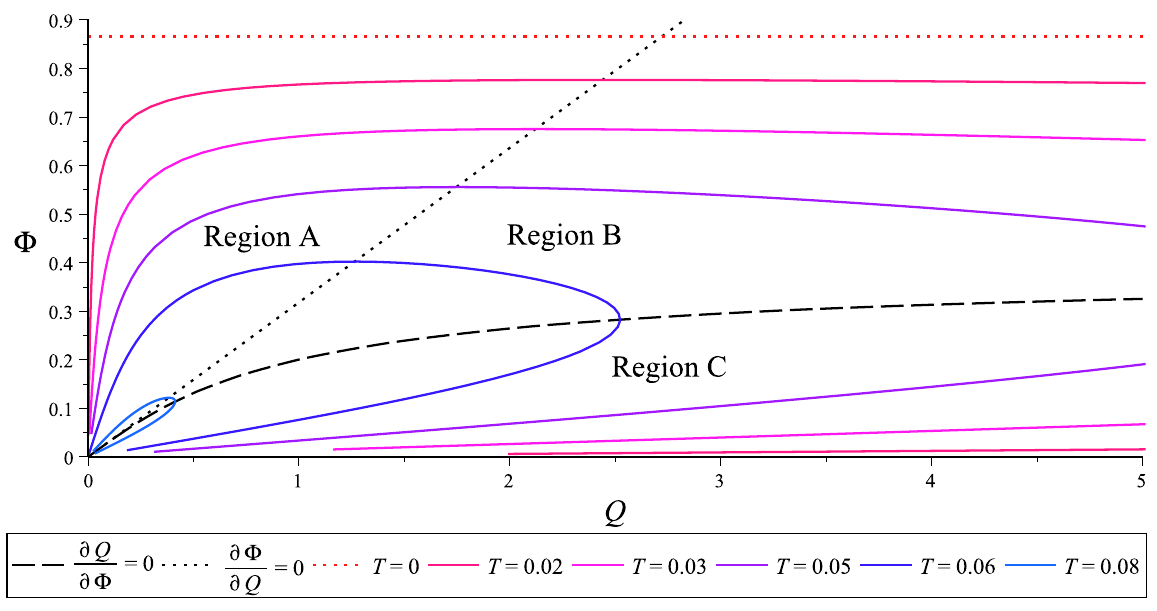}\;
	\includegraphics[width=0.5\textwidth]{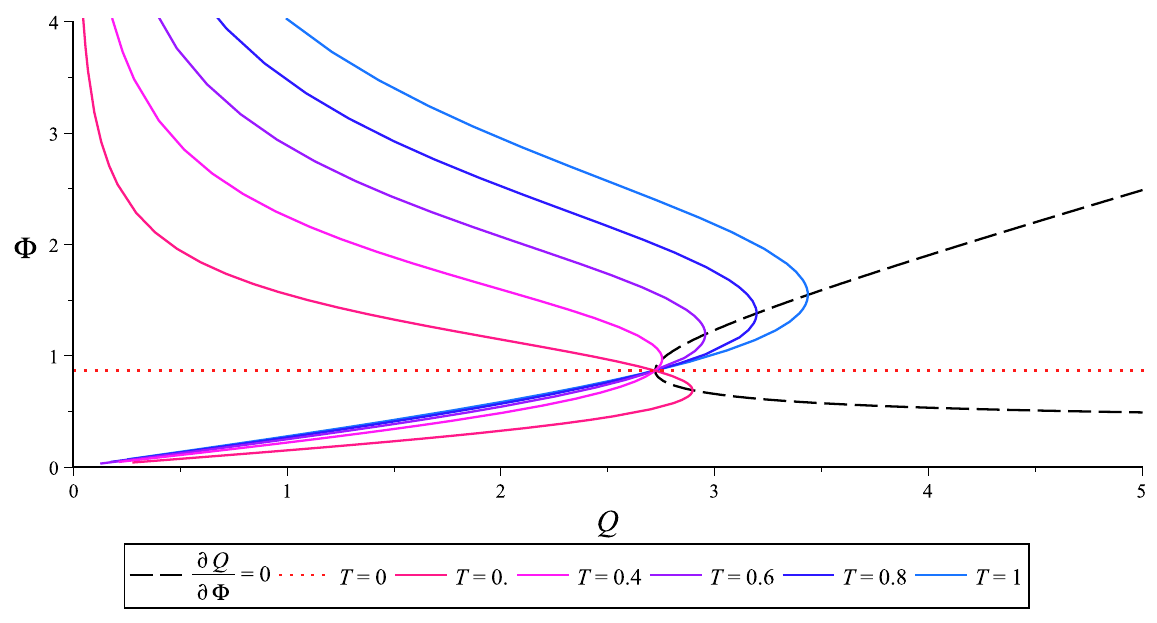}
	\caption{\small Equation of state $\Phi-Q$, for fixed $T$. \textbf{Left hand side:} $\alpha=1$. \textbf{Right hand side:} $\alpha=-1$.}
	\label{EoS1}
\end{figure}

For the family $\alpha>0$, the isotherms $T\neq 0$ in the $Q-\Phi$ plane are characterized by being `closed', i.e, starting and ending at $Q=\Phi=0$, as shown in the first plot of Fig. \ref{EoS1}. This indicates that there is one curve, say $\Phi_0=\Phi_0(Q)$, along which $\epsilon_T=0$, and another curve, say $\Phi_\infty=\Phi_\infty(Q)$, along which $\epsilon_T$ diverges or, alternatively, $\epsilon_T^{-1}=0$. From the expression of the isothermal permittivity,
\begin{equation}
	\epsilon_T=\frac{Q\(12\pi\alpha\Phi^3+20Q\Phi^2+3\pi\alpha\Phi-3Q\)} {(4\Phi^2-3)(Q-\pi\alpha\Phi)\Phi}
\end{equation}
it follows that the $\Phi_0=\Phi_0(Q)$ is given implicitly by the cubic equation
\begin{equation}
	\label{curve1}
	12\pi\alpha\Phi_0^3(Q)+20Q\Phi_0^2(Q)+3\pi\alpha\Phi_0(Q)-3Q=0
\end{equation}
whereas
\begin{equation}
	\label{curve2}
	\Phi_\infty(Q)=\frac{Q}{\pi\alpha}
\end{equation}
Because $\epsilon_T$ changes twice its sign along a given isotherm $T\neq 0$, we can distinguish three regions. Regions A and C contain electrically stable configurations ($\epsilon_T>0$), and region B contains electrically unstable ones. While region B and C mimic the behaviour for charged black holes in Einstein-Maxwell gravity, the presence of the GB invariant, moduled by the constant $\alpha$, is responsible for new stable configurations within region A, which consist of small black holes characterized by $\Phi>\frac{Q}{\pi\alpha}$, which is equivalent to $r_+^2<\alpha$.

Moreover, for $\alpha>0$, there exists a maximum allowed temperature that depends only on $\alpha$. By looking at the first plot in Fig. \ref{EoS1}, we can observe that, as temperature approaches to its maximum, $T_{max}$, the corresponding isotherm shrinks to eventually disappears. It can be shown, but also follows from visual inspection, that in the limit $T\rightarrow T_{max}$, $\Phi\rightarrow Q/(\pi\alpha)$. In other words, the isotherms tend to align with the dotted curve in Fig. \ref{EoS1}. By replacing $\Phi=Q/(\pi\alpha)$ in the equation of state, then, taking the limit $Q\rightarrow 0$ leads to 
\begin{equation}
	T_{max}=\frac{1}{4\pi\sqrt{\alpha}}	
\end{equation}
On the other hand, for $\alpha<0$, there is one special curve in the $Q-\Phi$ phase space, namely, $\Phi_0=\Phi_0(Q)$, where $\epsilon_T=0$. This situation is depicted in the second plot of Fig. \ref{EoS1}. As shown next, for the $\alpha<0$ case, the region with $\epsilon_T>0$ is thermally unstable, $C_Q<0$. 

\subsection{Phase diagram and local stability}

\subsubsection{Canonical ensemble: $Q$ fixed}

\begin{figure}[t!]
	\includegraphics[width=0.5\textwidth]{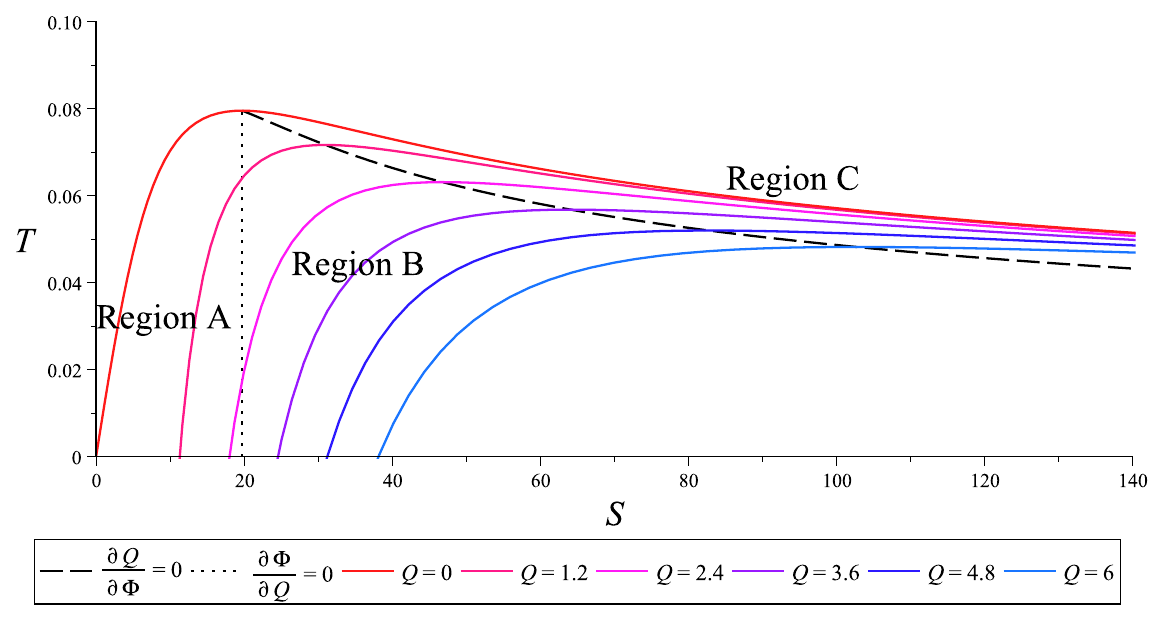}
	\includegraphics[width=0.5\textwidth]{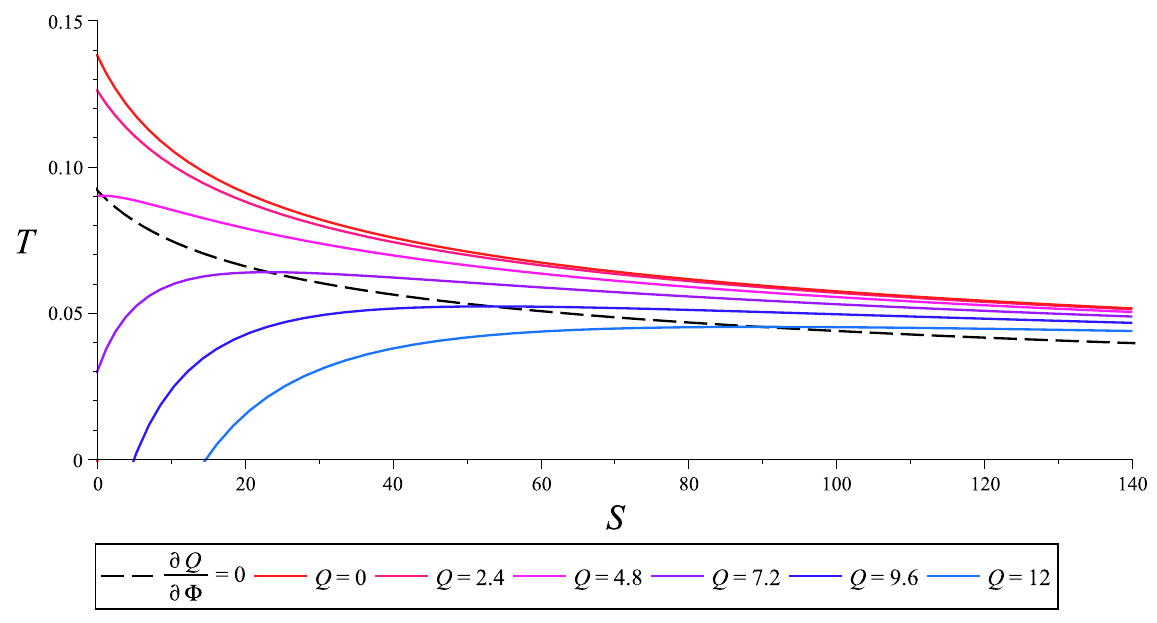}
	\caption{\small $T-S$ for fixed $Q$. \textbf{Left hand side:} $\alpha=1$. \textbf{Right hand side:} $\alpha=-1$.}
	\label{TS0}
\end{figure}

Let us focus on the case $\alpha>0$. From the equation of state, we have that, for a given $Q$ within the region A, the electrically stable configuration is the one that has the largest value of $\Phi$. We now explicitly show that configurations within region A are also thermally stable. The heat capacity in this case is
\begin{equation}
	C_Q=T\(\frac{\pa S}{\pa T}\)_Q=\frac{3\sqrt{\pi Q\Phi}(Q+\pi\alpha\Phi)^2\(3-4\Phi^2\)}{2\Phi^2(12\pi\alpha\Phi^3+20Q\Phi^2+3\pi\alpha\Phi-3Q)}
	\label{heat}
\end{equation}
It is easy to translate the curves that divide region A, B and C, given by (\ref{curve1}) and (\ref{curve2}), into the diagrams $T-S$, as shown in the first plot in Fig. \ref{TS0}. As commented earlier, region A contains small black holes, $r_+^2<\alpha$, that is $S<2\pi^2\alpha^\frac{3}{2}$. These configurations are thermally stable as $C_Q>0$, and thus, are locally thermodynamically stable.

\subsubsection{Grand canonical ensemble: $\Phi$ fixed}

The thermodynamic stability in the grand canonical ensemble can be investigated in the same way, though the analysis is simpler. The heat capacity for this ensemble is
\begin{equation}
	C_\Phi=\frac{3\sqrt{\pi Q}(Q+\pi\alpha\Phi)^2} {2\Phi^{\frac{3}{2}}(\pi\alpha\Phi-Q)}
	\label{heat2}
\end{equation}
Consider the case of interest, $\alpha>0$. Since region A is characterized by $\Phi>Q/(\pi\alpha)$, it immediately follows that $C_\Phi>0$ within this region. 

Thus, we conclude that region A contains fully thermodynamically stable configuration both in the canonical and grand canonical ensembles.

\section{Discussion}
\label{disc}
In this paper, we provide examples of thermodynamically stable asymptotically flat black holes. We have shown that effective theories with the dilaton and its self-interaction, as well as gravity with GB corrections, can allow for thermodynamic stability regions in the phase diagram of 
charged static black holes.

The conserved charges of asymptotically flat black holes are usually computed by using
Hamiltonian methods \cite{Arnowitt:1960es,Arnowitt:1960zzc,Arnowitt:1961zz,Arnowitt:1962hi}. However, supplemented with counterterms, the quasilocal formalism of Brown and York \cite{Brown:1992br} becomes a powerful framework for computing conserved quantities in general relativity. The basic idea in \cite{Brown:1992br} is to define a ‘quasilocal’ energy inside a given finite region that can be directly derived from the gravitational action for that specific spatially bounded region. The quasilocal energy is the value of Hamiltonian which generates unit magnitude
proper-time translations in a timelike direction orthogonal to spacelike hypersurfaces at
some fixed spatial boundary and so  it agrees with the ADM energy in the limit when the
spatial boundary is pushed to infinity. We emphasize, though, that while within the ADM formalism the foliation is made by hypersurfaces that are  Cauchy surfaces such that the data on a slice determine completely the future evolution of the system, that is not necessary the
case for the quasilocal formalism. One important result of our work was to obtain a consistent counterterm for GB gravity in flat spacetime. While the variational principle is at the basis of this construction, a relevant subtlety we had to deal with was the existence of a finite contribution coming from the counterterm that is related to the ambiguity in defining its overall factor. To fix this ambiguity and construct the general counterterm that regularizes the action  we had to rely on the Hamiltonian formalism. Once the quasilocal formalism is consistently supplemented with counterterms, it can be also used for theories with higher derivative corrections. We emphasize that, unlike in the case of Wald formalism, we do not need to use the first law a priori and so the quasilocal formalism is self-consistent providing all information about the thermodynamic behaviour of the gravitational system.

The existence of asymptotically flat hairy black holes in theories with a scalar field potential was proposed in \cite{Nucamendi:1995ex}. We made this proposal concrete by analyzing exact solutions when the dilaton together with its potential are present in the theory. Since in flat spacetime, the effective potential (obtained from the self-interaction of dilaton together with the non-trvial coupling to the gauge field) plays the role of the local ‘box’, one can expect that there can exist thermodynamcally stable black holes in such theories. This is indeed true  for 4-dimensional supergravity theory with  FI terms 
for which, interestingly, there exist exact hairy solutions. Unlike the black holes in AdS spacetime, the small black holes (the parameter $\Upsilon$ coming from the FI sector provide a scale in the theory) are stable. This can be understood as follows: when the horizon radius is
large, the dilaton potential gets weaker (it vanishes at the boundary) and so the large black holes are not stable, while for small ones, the 
self-interaction becomes relevant acting like a box allowing configurations in stable thermal equilibrium. There exists a family of solutions that contains two distinct branches, but only one branch contains thermodynamically stable hairy black holes (see the first plot in Fig. \ref{EoS3} and the first plot in Fig. \ref{EoS5}, where the stable black holes reside within region A, for the positive branch). The range for which these black holes exist is close to the extremality.

Surprisingly, the charged black holes in a gravity theory with GB corrections have a similar thermodynamic behaviour. As discussed in Section \ref{sec3}, there is again a family of asymptotically flat black holes with two different branches,  but only one branch contains thermodynamically stable black holes. In this case, it seems that the GB term in the action behaves as an `effective potential'. This can be explicitly checked by comparing the behaviour in the asymptotic region of $V(\phi)$ vs GB term:
\begin{equation}
	V(\phi)={\frac{\Upsilon}{30}}{\phi}^{5} 
	+{\frac{\Upsilon}{630}}{\phi}^{7}+\mathcal{O}\left({\phi}^{9}\right) =\frac{\Upsilon}{30\eta^5r^5}-\frac{3\Upsilon}{560\eta^7r^7}+\mathcal{O}(r^{-9})
\end{equation}
and
\begin{equation}
	\mathcal{R}_{GB}=R^2-4R_{\mu\nu}R^{\mu\nu}+R_{\mu\nu\sigma\rho}R^{\mu\nu\sigma\rho}=\frac{72\mu^2}{r^8}-\frac{360\mu q^2}{r^{10}}-\frac{336\(\mu^3\alpha+q^4\)}{r^{12}}+ \mathcal{O}\left(r^{-14}\right)
\end{equation}
We observe the rapid fall off that explains, in both cases, the existence of asymptotically flat solutions. However, in both cases, the backreaction deep in the bulk becomes relevant.

For GB theory, the small black holes in the range close to the extremality are again thermodynamically stable. However, there is now an extra constraint, namely the existence of a critical charge: only those with $Q<\frac{1}{2}\sqrt{3}\pi\alpha$ or, equivalently, $S<2\pi^2\alpha^{3/2}$ are thermodynamically stable. This can be also understood from the first plot in Fig. \ref{EoS1} and Fig. \ref{TS0}, where we see that one zone that contains the near-extremal black holes (those close to the horizontal dotted curve) is within Region A (stable), but the other is in Region B (unstable).

To summarize this striking analogy, let us concretely emphasize which branches are relevant for the existence of  thermodynamically stable black hole configurations. For the Einstein-Maxwell-scalar theory, stability criteria are met when the scalar field potential has positive concavity and is bounded from below. This occurs for one of two  families, namely the $\Upsilon>0$ one, where $\Upsilon$ is the (global) parameter that controls the strength of the self-interaction. Within this family, only the branch for which the scalar field is positively defined and satisfies some particular boundary condition contains thermodynamically stable black hole solutions. For charged black holes in the GB theory, stability criteria are met when the metric is asymptotically flat. This occurs for one of the two families, namely the one defined by $\epsilon=-1$, where $\epsilon=\pm 1$ defines the asymptotic structure of the metric, and, within this family, for one of the two branches, namely the one where $\alpha>0$, where $\alpha$ controls the strength of the GB correction in the action.

\begin{figure}[t!]
	\includegraphics[width=0.47\textwidth]{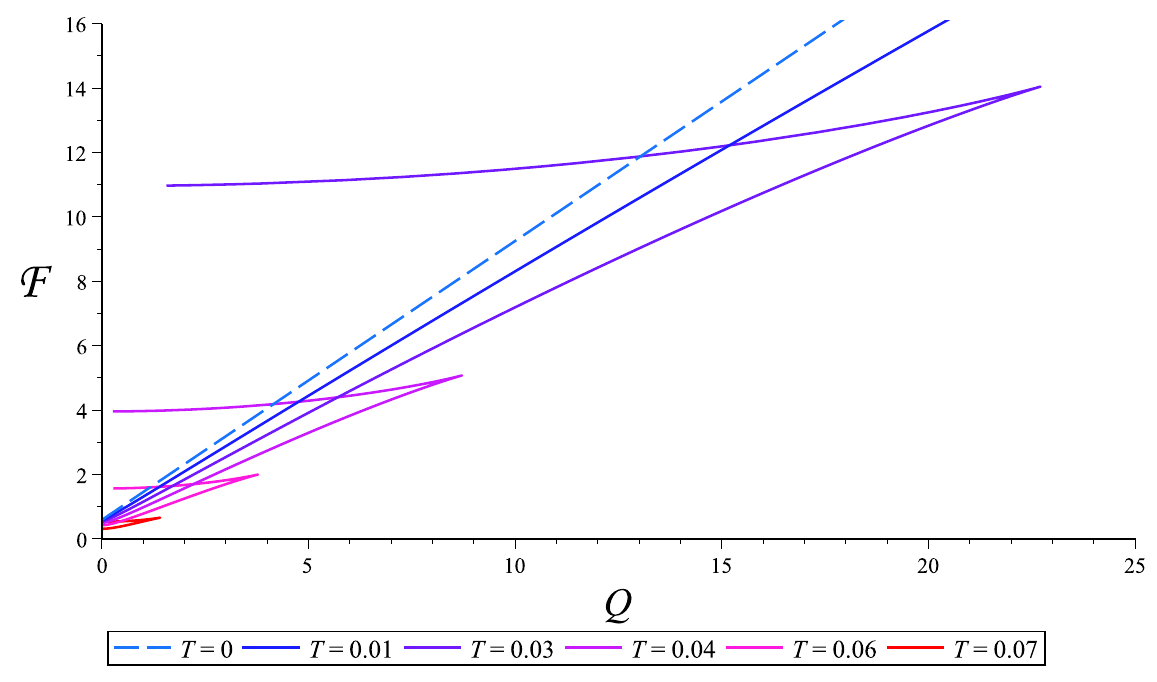}
	\includegraphics[width=0.49\textwidth]{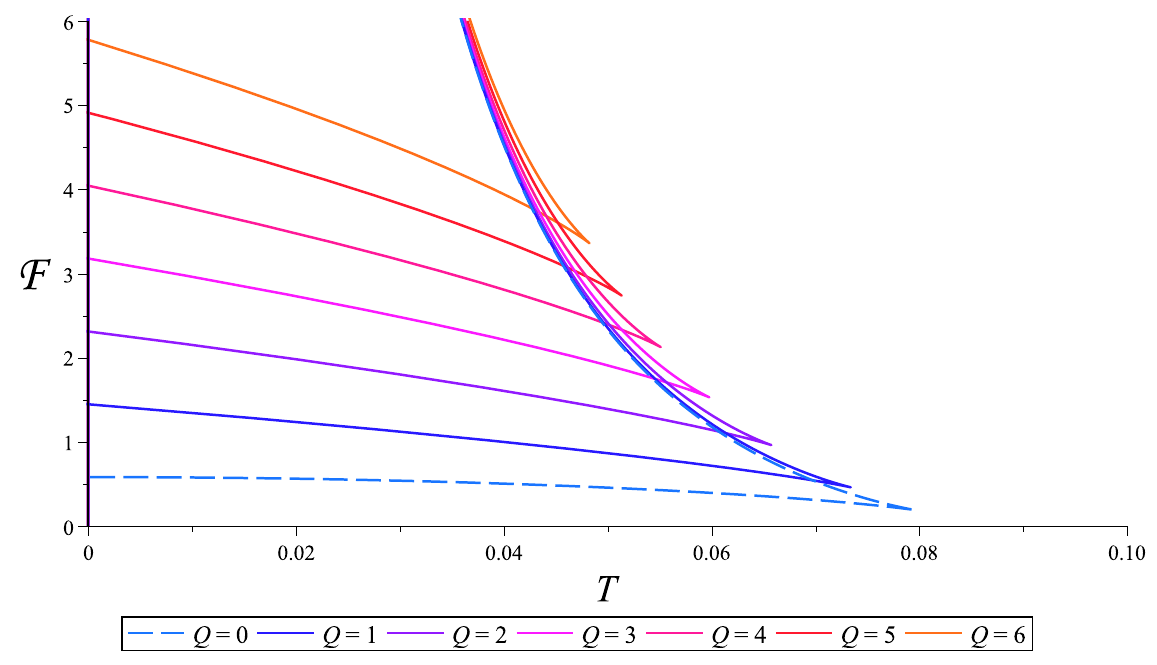}  
	\caption{\small \textbf{Left hand side:} $\mathcal{F}-Q$ for fixed $T$ for $\alpha=1$. \textbf{Right hand side:} $\mathcal{F}-T$ for fixed $Q$ for $\alpha=1$.}
	\label{FigFQ}
\end{figure}

To complement the thermodynamic analysis in the bulk of the paper that was based on the equation of state, we now briefly present an analysis of the thermodynamic potential. We start with the GB case and focus on the case of interest, namely $\alpha>0$. The relevant information obtained from the equation of state can also be read off from the  $\mathcal{F}-Q$ diagram when $T$ is fixed, depicted in the first plot of Fig. \ref{FigFQ}. The locally thermodynamically stable black holes are the ones in region A, i.e., those with the biggest value of $\Phi$, for a given $Q<\frac{1}{2}\sqrt{3}\pi\alpha$. Since $\Phi=\(\pa\mathcal{F}/\pa Q\)_T$, the stable configurations in $\mathcal{F}-Q$ diagram correspond to the lowest part of each isotherm, having the biggest slope. Also from Fig. \ref{FigFQ}, it follows that these configurations minimize the thermodynamic potential. From this observation, we conclude that locally stable black holes are also globally stable.
We shall investigate the  global stability using the $\mathcal{F}-T$ diagram, depicted in the second plot of Fig. \ref{FigFQ}. Since $C_Q=-T(\pa^2\mathcal{F}/\pa{T^2})_Q$, it follows that locally stable configurations are those satisfying $(\pa^2\mathcal{F}/\pa{T^2})_Q<0$, which also corresponds to the lower part of the curve, containing again configurations that minimize $\mathcal{F}$. 

\begin{figure}[t!]	
	\centering	\includegraphics[width=0.44\textwidth]{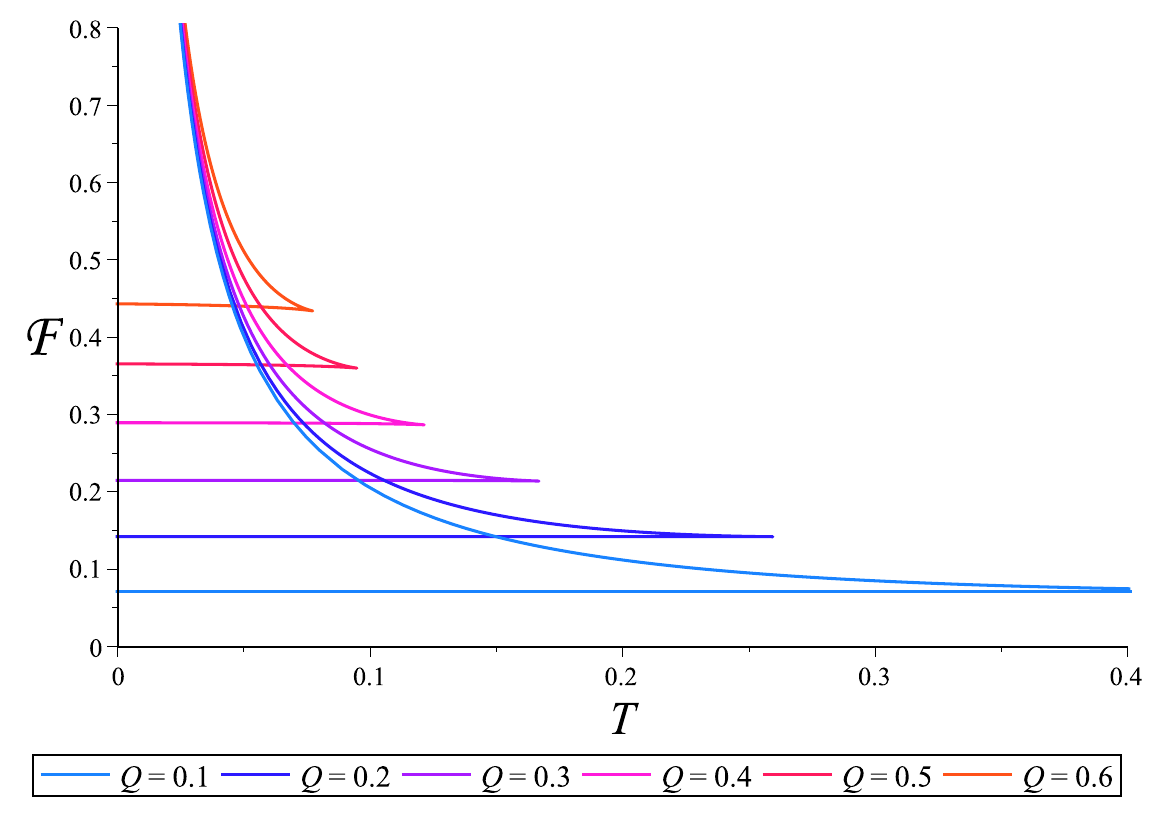}
	\caption{\small $\mathcal{F}-T$ for fixed $Q$, in the positive branch of the hairy solutions.}
	\label{TS1}
\end{figure}

The thermodynamic potential for the hairy black hole solution is depicted in Fig. \ref{TS1} where we consider the positive branch. However, the thermodynamically stable configurations are present only in the positive branch, corresponding to the first plot in Fig. \ref{TS1}. Since $C_Q=-T(\pa^2\mathcal{F}/\pa{T^2})_Q$, we observe that the locally stable configurations $C_Q>0$ are those with negative concavity, which correspond to the configurations that minimizes $\mathcal{F}$. 
Therefore, in a similar manner as we did in the GB case, we obtain that these solutions are also globally stable.

Since there is no solution of flat spacetime with constant charge, and unlike AdS spacetime where hairy charged solitons were explicitly constructed in some specific cases \cite{Anabalon:2021tua,Anabalon:2022aig,Anabalon:2023oge}, making sense of the ground state in this case is not obvious. However, as in \cite{Chamblin:1999tk}, one can consider the extremal black hole (otherwise the spacetime can collapse in a naked singularity) as the state with respect to which we compute the energy. Therefore, to complete the analysis, let us consider the extremal limit that is relevant for the existence of the canonical ensemble. One can use the entropy function formalism \cite{Astefanesei:2006dd,Sen:2005wa,Sen:2005iz} to obtain the near horizon geometry of black hole solutions in theories that are diffeomorphism and gauge invariant. This method is based on the
existence of an enhanced $AdS_2$ in the near horizon geometry that also plays  an important role for understanding the entropy of spinning \cite{Astefanesei:2006dd}  astrophysical black holes \cite{Guica:2008mu}. For asymptotically flat hairy black holes in a theory with a dilaton potential, this analysis was done 
in \cite{Astefanesei:2019qsg} (see, also, \cite{Astefanesei:2019pfq} for a related discussion in a different context) and for GB charged hairy black holes, the details can be found in \cite{Astefanesei:2008wz} and so we do not repeat them here. We would only like to emphasize that, for the extremal black holes we have considered in our work, the entropy does not vanish and so these are regular solutions of the equations of motion. In the GB case, the results can be obtained analytically, as follows: consider $T=0$ limit, namely $r_+^2=q$, as follows from (\ref{temper}). By replacing this value in the horizon equation, $f(r_+)=0$ or, equivalently, in (\ref{horizon2}), we get the following relation between the conserved charges for the extremal black hole: 
\begin{equation}
	E=\frac{1}{16}\({3\pi\alpha}+8\sqrt{3}Q\)
\end{equation}
and the entropy becomes
\begin{equation}
	S
	=\frac{(2\pi Q)^\frac{1}{2}}{2\cdot 3^\frac{1}{4}}\(3\pi\alpha+\frac{2Q}{\sqrt{3}}\)
\end{equation}

\begin{figure}[t!]\centering	\includegraphics[width=0.74\textwidth]{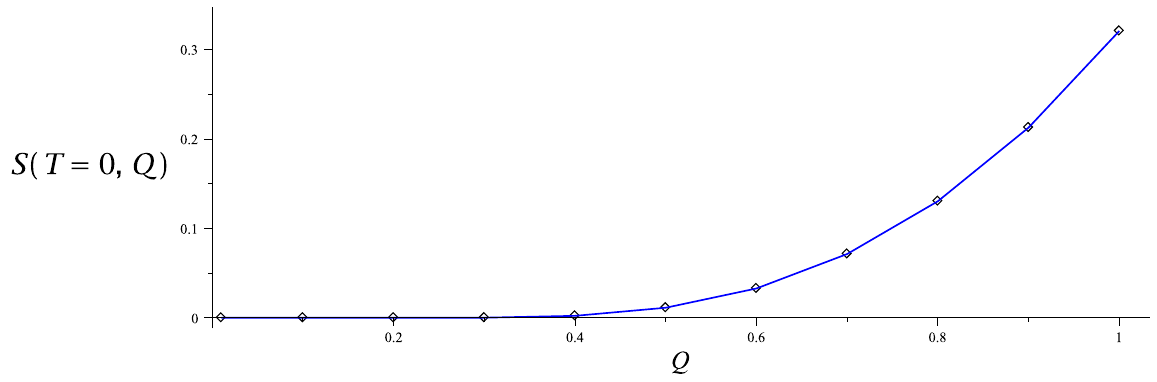}
	\caption{\small Entropy of extremal black hole for the hairy black holes in the positive branch. $\Upsilon=1$.}\label{extremalhairy}
\end{figure}

In the hairy case,  we can show, through numerical computations, that the entropy of extremal black holes is also positive, with its values depending of the electric charge. It turns out that, for increasing values of $Q$, the entropy of extremal hairy black hole rapidly approaches to zero, as depicted in Fig. \ref{extremalhairy}. This can be also seen from the first plot of Fig. \ref{EoS5}, though, from that plot and for curves with large values of $Q$, it is not as easy to discern that $S(T=0)>0$.

While the theories are completely different, we would like to point out that, generally, theories with higher derivatives $f(R)$ are equivalent with theories with a scalar field and its self-interaction \cite{Sotiriou:2006hs, Sotiriou:2008rp}. However, this consideration can not be applied for our work because one theory is $4$-dimensional, while the other is $5$-dimensional. The GB term is trivial in $4$-dimensions, but once it is coupled with the scalar field it contributes to the equations of motion and so we expect that the hairy black holes receive corrections. Since we were not able to generate exact solutions, we leave the detailed analysis of this specific case for a future work.

\section{Acknowledgments}
	The work of DA, PC, and GC was supported by the Fondecyt Regular Grant 1242043. The work of RB has been supported by the
	National Agency for Research and Development [ANID] Chile, Doctorado Nacional under grant 2021-21211461. 
	RR thanks the financial support from ANID, by means of FONDECYT Postdoctorado grant number 3220663.

\newpage
\appendix

\section{Local stability conditions}
\label{localconditions}

The local stability of equilibrium configurations follows from demanding that heat capacity $C_Q\equiv T(\pa S/\pa T)_Q$ and isothermal permittivity $\epsilon_T\equiv (\pa{Q}/\pa\Phi)_T$ to be simultaneously positive defined. The proof follows from considering that at stable equilibrium, the entropy has a maximum value with respect to the entropy of the system when considering the fluctuations. We can use the argument (see, for instance, \cite{callen}) that, if $S(E,Q)$ represents the entropy before some fluctuations in $E$ and $Q$, given by $\delta E$ and $\delta Q$, respectively, then, for the configuration $S(E,Q)$ to be stable, the average entropy $\frac{1}{2}\[S(E+\delta E,Q+\delta Q)+S(E-\delta E,Q-\delta Q)\]$ after the perturbation cannot be greater than the initial one. The same argument is valid in terms of the energy: for a given state $E=E(S,Q)$ to be locally stable, the average energy after the perturbation can not be lower than the initial one. This gives rise to three mathematical conditions for local stability, namely,
\begin{equation}
	\label{conditions}
	\(\frac{\pa^2E}{\pa S^2}\)_{Q}\(\frac{\pa^2E}{\pa Q^2}\)_{S}-\[\(\frac{\pa}{\pa S}\)_Q\(\frac{\pa E}{\pa Q}\)_{S}\]^2>0\,, \qquad \(\frac{\pa^2E}{\pa S^2}\)_{Q}>0\,, \qquad 	\(\frac{\pa^2E}{\pa Q^2}\)_{S}>0
\end{equation}

Consider the canonical ensemble, for which the thermodynamic potential is
$\mathcal{F}=E-TS$. For infinitesimal changes in its thermodynamics, we can describe the system by 
$d\mathcal{F}(T,Q)=-S d T+\Phi d Q$, and it can be shown, using the standard thermodynamic relations, that the conditions (\ref{conditions}) can be written in terms of $\mathcal{F}$ and, indeed, can be reduced to only two independent requirements:
\begin{align}
	C_Q&=T\(\frac{\pa S}{\pa T}\)_Q=-T\(\frac{\pa^2\mathcal{F}}{\pa{T}^2}\)_Q>0 \\
	\epsilon_T&=\(\frac{\pa Q}{\pa \Phi}\)_T =\(\frac{\pa^2\mathcal{F}}{\pa{Q^2}}\)_T^{-1}>0
\end{align}
where we have used $S=-\({\pa\mathcal{F}}/{\pa{T}}\)_Q$ and $\Phi=\({\pa\mathcal{F}}/{\pa{Q}}\)_T$.

In the grand canonical ensemble, it can be shown that stability also follows from analyzing the concavity of the thermodynamic potential $\mathcal{G}$. In this case, $\mathcal{G}=E-TS-\Phi Q$, the first law takes the form $d\mathcal{G}(T,\Phi)=-S d T-Q d \Phi$, and the relevant response function for local stability in this case is
\begin{equation}
	C_\Phi\equiv T\(\frac{\pa S}{\pa T}\)_\Phi =-T\(\frac{\pa^2\mathcal{G}}{\pa{T}^2}\)_\Phi>0 
\end{equation}
where $S=-\({\pa\mathcal{G}}/{\pa{T}}\)_\Phi$.

\section{Dilaton counterterms in AdS vs flat spacetime}
\label{SC}

Let us consider the Einstein's equation derived from Einstein-Maxwell-scalar gravity (\ref{action2}), but 
now with the general potential (see, e.g., \cite{Anabalon:2013sra} where the extremal limit was also discussed)
\begin{equation}
	V(\phi)=\frac{2\Lambda}{3}\(2+\cosh\phi\)+2\Upsilon\(2\phi+\phi\cosh\phi-3\sinh\phi\)
\end{equation}
that contains the comological constant $\Lambda$. We now consider the asymptotically AdS solution, for which the limit $\Lambda \rightarrow 0$ leads to the asymptotically flat solution,
\begin{equation}
	ds^2=\Omega(x)\[-f(x)dt^2+\frac{\eta^2dx^2}{x^2f(x)}+d\theta^2+\sin^2\theta d\phi^2\]\,, \quad A_\mu=\[\Phi-\frac{q(x-1)}{x}\]\delta_\mu^t
\end{equation}
where
\begin{equation}
	\Omega(x)=\frac{x}{\eta^2(x-1)^2}\,, \qquad f(x)=-\frac{\Lambda}{3}+\Upsilon\, \left({\frac{{x}^{2}-1}{2x}}-\ln{x}\right) +{\frac{{\eta}^{2}\left(x-1\right)^{2}}{x} \left[1-{\frac {2\left( x-1 \right) {q}^{2}}{x}}
		\right] }
\end{equation}
By taking the trace of the Einstein's equation, we can use the relation
$R=\frac{1}{2}(\pa\phi)^2+2V$
to simplify the bulk part of the action that yields
\begin{equation}
	I_{bulk}=\frac{1}{2\kappa}\int_{\mathcal{M}}
	{d^4x\sqrt{-g}\[R -\frac{1}{2}(\pa\phi)^2- e^\phi{F^2}-V(\phi)\]}
	=\frac{1}{2\kappa}\int_{\mathcal{M}}
	{d^4x\sqrt{-g}\[V(\phi)-e^\phi{F^2}\]}
\end{equation}
The on-shell Euclidean action is
\begin{equation}
	I_{bulk}^E=\frac{1}{4}\beta\int_{x_+}^{x_b}dx\[\frac{2\eta\Omega(x)}{x}-\frac{(xf(x)\Omega'(x))'}{\eta}\]
\end{equation}
Together, the bulk part and Gibbons-Hawking boundary term add up to
\begin{equation}
	I_{bulk}^E+I_{GH}^E
	=\frac{\beta\[(12\eta^2q^2-\Upsilon)(x_+-1)-6\eta^2(x_+-3)\]}{24\eta^3(x_+-1)} -\frac{\beta\(6\eta^2-\Lambda\)}{6\eta^3(x_b-1)} +\frac{\beta\Lambda}{2\eta^3(x_b-1)^2} +\frac{\beta\Lambda}{3\eta^3(x_b-1)^3}
\end{equation}
where $x_b \rightarrow 1$ is the boundary location. We note that there are three divergent terms proportional with $\Lambda$ that are not going to survive for the asymptotically flat solution when the cosmological constant vanishes. The gravitational counterterm required to remove the divergences is specific to the flat and AdS spacetimes. For asymptotically flat spacetime, the gravitational counterterm is
\begin{equation}
	I_{ct(flat)}^E=\frac{1}{\kappa}\int_{\pa\mathcal{M}}{d^3x\sqrt{|h|}\sqrt{2\mathcal{R}^{(3)}}}=\frac{\beta}{\eta(x_b-1)}-\frac{\beta\(12\eta^2q^2-6\eta^2-\Upsilon\)}{12\eta^3}+\mathcal{O}(x_b-1)
\end{equation}
It is clear that the counterterm for asymptotically flat spacetime perfectly cancels the only divergence coming from $I_{bulk}+I_{GH}$ when $\Lambda=0$. The total action for flat spacetime satisfies the quantum-statistical relation $\beta^{-1}I^E=E-TS-\Phi Q$, as shown in Section \ref{sec2}. 

Now, the gravitational counterterm for asymptotically AdS spacetime is
\begin{align}
	I_{ct,(AdS)}^E&=\frac{1}{\kappa}\int_{\pa\mathcal{M}}{d^3x\sqrt{h}\[\frac{2}{\ell}+\frac{\ell}{2}\mathcal{R}^{(3)}\]} \\ &=\frac{\beta\(\Lambda+4\Upsilon+24\eta^2-48\eta^2q^2\)}{48\eta^3} +\frac{\beta(8\eta^2-\Lambda)}{8\eta^3(x_b-1)} -\frac{\beta\Lambda}{2\eta^3(x_b-1)^2} -\frac{\beta\Lambda}{3\eta^3(x_b-1)^3}
	\label{counter2}
\end{align}
where $\Lambda=-3/\ell^2$.
For the AdS case, the divergence coming from $I_{bulk}+I_{GH}$ are not completely cancelled by (\ref{counter2}), and one of the terms $\propto(x_b-1)^{-1}$ survives. Concretely,
\begin{equation}
	I_{bulk}^E+I_{GH}^E+I_{ct,(AdS)}^E =\frac{\beta\[12{\eta}^{2}(x_++1) +(\Lambda+2\Upsilon-24{\eta}^{2}{q}^{2})(x_+-1)\]}{48\eta^3(x_+-1)} +\frac{\beta\Lambda}{24\eta^3(x_b-1)}
\end{equation}
and the remaining divergent term is cancelled by the contribution from the scalar field counterterm \cite{Marolf:2006nd, Anabalon:2015xvl} (consistent with the Hamiltonian formalism \cite{Hertog:2004ns, Anabalon:2014fla})
\begin{equation}
	I_{\phi}^E=\frac{1}{2\kappa}\int_{\pa\mathcal{M}}{d^3x\sqrt{h}\[\frac{\phi^2}{2\ell}+\frac{W(\phi)}{\ell A}\phi^3\]}=-\frac{\beta\Lambda}{24\eta^3(x_b-1)}-\frac{\beta\Lambda}{48\eta^3}
\end{equation}
where $\phi(r)=A/r+B/r^2+\cdots$, $r=\sqrt{\Omega(x)}$, and $B=dW(A)/dB$ (in this case, $B=0$ and $W=0$). The total action also satisfies the quantum-statistical relation, $\beta^{-1}I^E=E-TS-\Phi Q$, where $I^E=I^E_{bulk}+I_{GH}^E+I_{ct,(AdS)}^E+I^E_{\phi}$.


\end{document}